\documentclass[journal]{IEEEtran}
 \usepackage{graphicx,color}
\ifCLASSINFOpdf
\else
\fi
\begin{document}
\title{Blockchain and Distributed Ledger Technologies for Cyberthreat Intelligence Sharing }

\author{Asadullah Tariq,~\IEEEmembership{Member IEEE}, Tariq Qayyum,~\IEEEmembership{Member IEEE}, Saed Alrabaee,~\IEEEmembership{Senior Member IEEE},, Mohamed Adel Serhani
        % <-this % stops a space
\thanks{A.Tariq, T.Qayyum and S.Alrabae are with CIT, UAEU,
GA, 30332 USA e-mail: (700039114@uaeu.ac.ae, 700036923@uaeu.ac.ae, salrabaee@uaeu.ac.ae). M.A.Serhani is with University of Sharjah (Email: Mserhani@sharjah.ac.ae)}% <-this % stops a space
\thanks{Asadullah is with United Arab Emirates University.}% <-this % stops a space
\thanks{}}
\markboth{}%
{}
\maketitle

\begin{abstract}
Cyberthreat intelligence sharing is a critical aspect of cybersecurity, and it is essential to understand its definition, objectives, benefits, and impact on society. Blockchain and Distributed Ledger Technology (DLT) are emerging technologies that have the potential to transform intelligence sharing. This paper aims to provide a comprehensive understanding of intelligence sharing and the role of blockchain and DLT in enhancing it. The paper addresses questions related to the definition, objectives, benefits, and impact of intelligence sharing and provides a review of the existing literature. Additionally, the paper explores the challenges associated with blockchain and DLT and their potential impact on security and privacy. The paper also discusses the use of DLT and blockchain in security and intelligence sharing and highlights the associated challenges and risks. Furthermore, the paper examines the potential impact of a National Cybersecurity Strategy on addressing cybersecurity risks. Finally, the paper explores the experimental set up required for implementing blockchain and DLT for intelligence sharing and discusses the curricular ramifications of intelligence sharing.
\end{abstract}


% \fi
\IEEEpeerreviewmaketitle
\section{Introduction}
\IEEEPARstart{I}{ntelligence} sharing is a critical aspect of cybersecurity, and it has become increasingly important in recent years due to the growing number of cyber threats. It involves the exchange of information between different organizations to help prevent and mitigate cyber attacks. The success of intelligence sharing depends on the accuracy, confidentiality, and timeliness of the information exchanged. However, intelligence sharing is not without its challenges, including the risk of data breaches, the lack of standardization, and the lack of trust between organizations. Blockchain and Distributed Ledger Technology (DLT) are emerging technologies that have the potential to transform intelligence sharing. Blockchain is a decentralized digital ledger that can securely record transactions, while DLT is a distributed database that can store and share information. These technologies have unique features that make them well-suited for intelligence sharing, such as immutability, transparency, and decentralization.

This paper aims to provide a comprehensive understanding of intelligence sharing and the role of blockchain and DLT in enhancing it. The paper will address questions related to the definition, objectives, benefits, and impact of intelligence sharing and provide a review of the existing literature. Additionally, the paper will explore the challenges associated with blockchain and DLT and their potential impact on security and privacy. The paper will also discuss the use of DLT and blockchain in security and intelligence sharing and highlight the associated challenges and risks. Furthermore, the paper will examine the potential impact of a National Cybersecurity Strategy on addressing cybersecurity risks. Finally, the paper will explore the experimental set up required for implementing blockchain and DLT for intelligence sharing and discuss the curricular ramifications of intelligence sharing. In summary, this paper aims to provide a comprehensive understanding of intelligence sharing and the role of blockchain and DLT in enhancing it. The paper will explore various questions related to intelligence sharing, blockchain, and DLT and highlight the potential benefits and challenges associated with their implementation. The paper will also discuss the experimental set up required for implementing blockchain and DLT for intelligence sharing and examine the curricular ramifications of intelligence sharing.

The organization of the paper is as follow: Section 2 provides an overview of intelligence sharing, including its definition, objectives, and benefits. Section 3 provides an overview of blockchain technology and Distributed Ledger Technology (DLT), including their applications, methodologies, and advantages and disadvantages. Section 4 delves into the challenges and risks associated with blockchain technology, including areas with good business fit, distributed ledger taxonomy, and challenges in enhancing security and privacy with DLT. Section 5 explores the role of Distributed Ledger Technology (DLT) and blockchain in intelligence sharing, focusing on how they can enhance intelligence sharing. Section 6 discusses traditional methods used for intelligence sharing and presents a comprehensive review of existing literature in this field. Section 7 analyzes the National Cybersecurity Strategy and its implications for addressing cybersecurity risks associated with intelligence sharing. Finally, Section 8 presents an experimental set up for implementing Blockchain and Distributed Ledger Technology (DLT) for intelligent sharing, including available datasets and appropriate metrics for measuring accuracy. By addressing these sections, the paper aims to provide a comprehensive understanding of intelligence sharing, blockchain technology, and their intersection, as well as potential solutions and challenges in enhancing intelligence sharing using DLT and blockchain technology.

\section{Overview of Intelligence Sharing}

Intelligence sharing is a critical aspect of modern society, playing a pivotal role in ensuring the security and resilience of nations, organizations, and individuals. The concept refers to the exchange of information and knowledge between different entities, such as countries, organizations, or individuals, to enhance decision-making, improve security, and facilitate collaborative efforts. The significance of intelligence sharing lies in its ability to foster trust, increase situational awareness, and enable proactive responses to emerging threats. As the world becomes increasingly interconnected and complex, the importance of effective intelligence sharing cannot be overstated \cite{1}.

One of the primary objectives of intelligence sharing is to enhance the decision-making capabilities of the involved parties by providing timely and accurate information. In addition, intelligence sharing facilitates collaboration between different entities, allowing them to pool resources and knowledge to address common threats and challenges. By improving the overall security posture and enabling proactive responses to emerging threats, intelligence sharing contributes to the development of best practices and strategies for mitigating risks. Intelligence sharing also has significant implications for countering terrorism, organized crime, and other illicit activities. By leveraging the collective resources and knowledge of various entities, intelligence sharing can strengthen efforts to identify, track, and disrupt these activities. Moreover, it supports international cooperation and diplomacy by enabling the exchange of information on mutual security concerns, thereby fostering trust and cooperation between nations. Another crucial aspect of intelligence sharing lies in its potential to improve cybersecurity measures. By sharing threat intelligence, vulnerabilities, and mitigation strategies, organizations can bolster their defenses against cyber attacks and enhance the overall resilience of critical infrastructures and assets. In turn, this helps protect sensitive data and systems from unauthorized access, theft, or destruction\cite{2}.

The lack of effective intelligence sharing can have severe consequences for individuals, organizations, and nations. For example, in the aftermath of the 9/11 terrorist attacks, it became evident that the inability to share intelligence among different government agencies had contributed to the failure to detect and prevent the attacks. Since then, intelligence sharing has become a top priority in national security efforts, with numerous initiatives and programs established to facilitate the exchange of information and knowledge across borders and sectors. The importance of intelligence sharing is further underscored by recent events, such as the rise of state-sponsored cyberattacks and the increasing sophistication of cybercriminals. These incidents highlight the need for greater collaboration and coordination among different entities to protect vital assets, infrastructure, and information\cite{3}.

The benefits of effective intelligence sharing are numerous, including the ability to provide a highly secure and trustworthy electronic identity. By enabling the exchange of verified, trusted information, intelligence sharing can help establish a reliable digital identity for individuals and organizations. This, in turn, contributes to increased trust and security in online transactions and interactions. Data confidentiality, integrity, and availability are also essential aspects of intelligence sharing. Confidentiality ensures that sensitive information is protected from unauthorized access, while integrity guarantees that the data remains accurate and consistent throughout its lifecycle. Availability refers to the accessibility of information when needed. By implementing robust security measures and protocols, intelligence sharing can help ensure that these critical aspects of information security are maintained. Privacy is another important consideration in intelligence sharing. As organizations and governments exchange sensitive information, it is crucial to protect the privacy of individuals and entities involved. Through the use of advanced encryption techniques and secure communication channels, intelligence sharing can balance the need for information exchange with the protection of personal privacy.

The severe impact of inadequate intelligence sharing can be seen through various examples and statistics. For instance, in the realm of cybersecurity, it has been estimated that cybercrime costs the global economy approximately \$6 trillion annually. Much of this damage could be mitigated or prevented through more effective sharing of threat intelligence and best practices among organizations and nations. The increasing number and scale of cyberattacks worldwide further emphasize the need for improved intelligence sharing to protect critical infrastructure, sensitive information, and global economic stability\cite{4}.

Another example that illustrates the importance of intelligence sharing is the WannaCry ransomware attack in 2017. The attack affected over 200,000 computers across 150 countries, causing widespread disruption and financial losses. WannaCry exploited a vulnerability in the Windows operating system, which had been discovered and subsequently disclosed by the United States National Security Agency (NSA) \cite{5}. However, the information about this vulnerability was not shared with relevant parties in a timely manner, allowing cybercriminals to take advantage of it before patches could be widely deployed. This incident underscores the need for effective intelligence sharing to prevent or mitigate the impact of cyberattacks\cite{6}. The role of intelligence sharing in thwarting terrorist attacks is also significant. For example, in 2015, intelligence agencies from France, Belgium, and other European countries collaborated to identify and apprehend the terrorists responsible for the Paris attacks, which resulted in the deaths of 130 people. The success of this operation highlights the value of intelligence sharing in identifying and neutralizing threats, ultimately saving lives and preserving national security \cite{7}, \cite{m1}.

In addition to these examples, various initiatives and organizations have been established to facilitate intelligence sharing on a global scale. Examples of such initiatives include the Five Eyes intelligence alliance, which comprises Australia, Canada, New Zealand, the United Kingdom, and the United States\cite{8}. This alliance enables member countries to share intelligence information and collaborate on joint operations, significantly enhancing their collective security capabilities. Similarly, organizations such as the European Union Agency for Law Enforcement Cooperation (Europol) and the International Criminal Police Organization (INTERPOL) facilitate intelligence sharing among their member countries to combat transnational crime and terrorism \cite{9}. Through these organizations, countries can pool resources and expertise to address complex, cross-border threats more effectively\cite{10}.

Despite the clear benefits of intelligence sharing, several challenges must be addressed to ensure its effectiveness. These challenges include the need for standardization of information formats and communication protocols, as well as the establishment of trust between participating entities. Additionally, concerns regarding privacy and data protection must be carefully balanced against the need for information exchange \cite{m2}.

Blockchain and distributed ledger technologies (DLT) have emerged as promising solutions to address these challenges. With their inherent characteristics of trust, transparency, and security, these technologies can facilitate seamless, secure, and privacy-preserving intelligence sharing. By leveraging blockchain and DLT, researchers and practitioners can develop innovative approaches to enhance security and resilience in an increasingly interconnected and complex world. In conclusion, intelligence sharing plays a critical role in ensuring the security and stability of nations, organizations, and individuals in an increasingly interconnected world. By fostering trust, enhancing situational awareness, and enabling proactive responses to emerging threats, intelligence sharing contributes to the development of best practices and strategies for mitigating risks. As the importance of effective intelligence sharing continues to grow, the potential of blockchain and distributed ledger technologies to address its challenges and unlock its full potential is an exciting area of exploration for researchers and practitioners alike.Blockchain technology based intelligence sharing is illustrated in Figure 1.

\begin{figure*}[!ht]
\centering
\includegraphics[width=15cm,height=13cm,keepaspectratio]{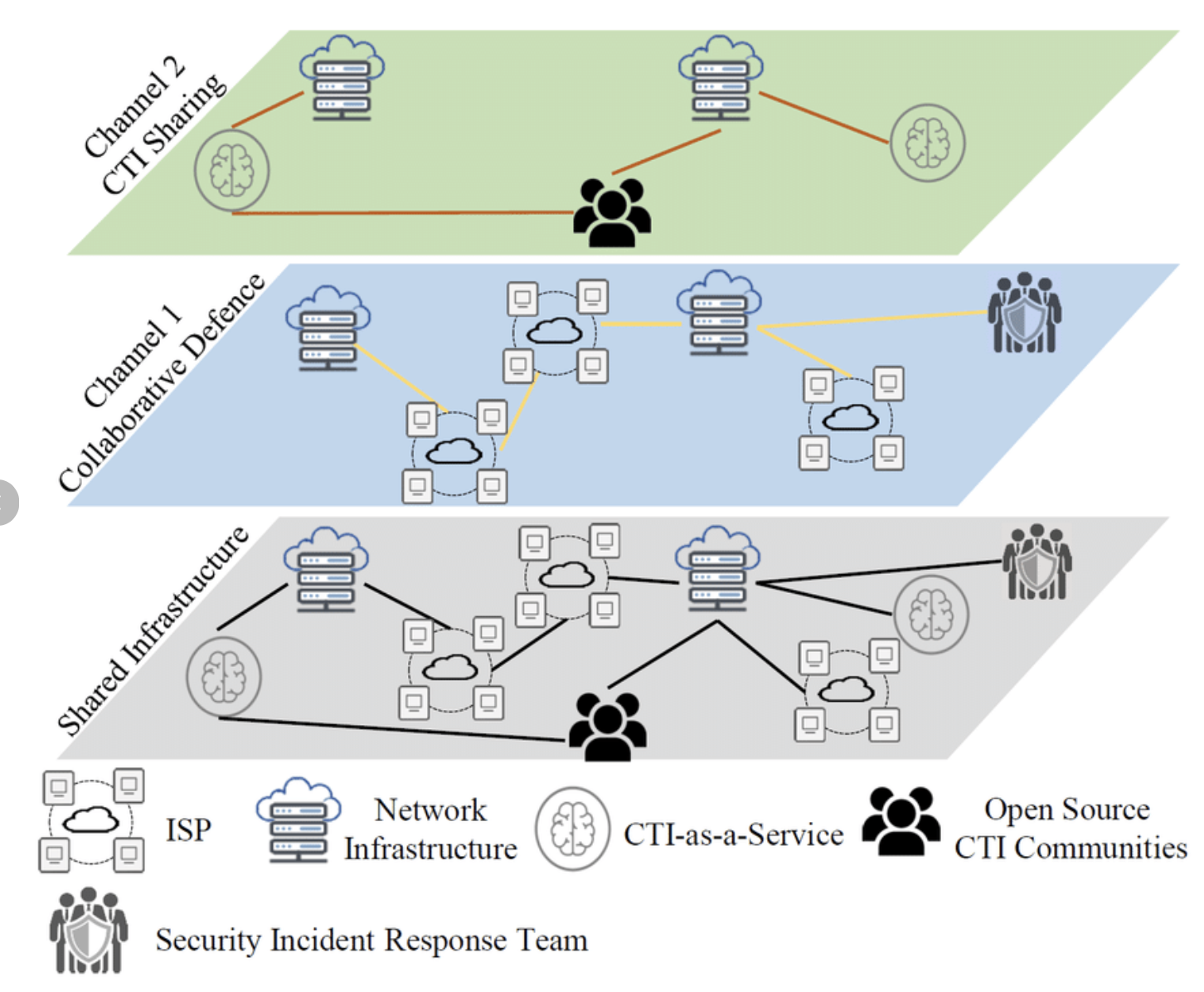}
\caption{Illustration of Cyber threat intelligence sharing for different scenarios using blockchain technology.}
\label{fog}
\end{figure*}
\section{Overview of Blockchain and Distributed Ledger Technology}

Blockchain technology and Distributed Ledger Technology (DLT) have emerged as transformative innovations that hold immense potential to revolutionize various industries. These decentralized, digital ledgers record transactions across multiple computers or nodes, ensuring that the data is secure and tamper-proof. By examining their applications, methodologies, advantages, and disadvantages, we can gain a comprehensive understanding of these technologies and their impact on the business world\cite{11}, \cite{m3}. Blockchain technology is a specific form of DLT that uses blocks linked together chronologically through cryptography, forming a secure chain of data. Distributed Ledger Technology, on the other hand, encompasses a broader range of decentralized database systems, of which blockchain is a subset. These technologies have found applications in numerous sectors, including finance, supply chain management, healthcare, and voting systems, to name just a few \cite{m4}. The transformative power of blockchain and DLT lies in their ability to enhance transparency, security, and efficiency in various processes, disrupting traditional business models and paving the way for innovative solutions\cite{12}.

Common methodologies for implementing blockchain and DLT involve the use of consensus algorithms, cryptographic techniques, and smart contracts. Consensus algorithms, such as Proof of Work (PoW) and Proof of Stake (PoS), ensure that all participating nodes in the network agree on the state of the ledger. These algorithms serve as the foundation for decentralized networks, allowing them to operate without a central authority. Cryptographic techniques, including public-key cryptography and hash functions, provide data security and integrity. These encryption methods protect sensitive information from unauthorized access, while also ensuring that the data cannot be tampered with once it is added to the ledger. Smart contracts are self-executing agreements encoded in the blockchain, which automate transactions and reduce the need for intermediaries \cite{13}. These programmable contracts enable complex transactions to be executed automatically, based on pre-defined conditions, increasing efficiency and trust among participants.

Understanding the inner workings of blockchain and DLT technologies is essential to appreciate their advantages and disadvantages. Both technologies offer several benefits, such as improved data security, transparency, and reduced reliance on intermediaries. They also enable faster, more efficient transactions and increased trust among participants \cite{m5}. However, these technologies also present challenges, including scalability, energy consumption, and regulatory concerns. Blockchain networks, particularly those using PoW consensus algorithms, are known to consume significant amounts of energy, raising sustainability concerns. Additionally, the decentralized nature of these technologies raises questions about regulatory oversight, as traditional centralized authorities struggle to adapt to the new paradigm \cite{14}.

The transition of blockchain technology from hype to reality has been driven by a growing recognition of its practical applications and the development of robust solutions addressing its limitations. Business initiatives focused on leveraging blockchain and DLT for improved efficiency, cost reduction, and enhanced security have contributed to the growth of these technologies. Major financial institutions are adopting DLT solutions to streamline cross-border payments, while supply chain companies are using blockchain technology for increased visibility and traceability of goods \cite{15}. Governments and public sector organizations are also exploring the use of blockchain and DLT for various applications, such as land registry management and digital identity systems. The scale and transformation of transactions in the decentralized digital age have been significant. With the growing adoption of blockchain and DLT, the number of daily transactions and the overall transaction volume have increased dramatically. These technologies have facilitated the creation of new digital assets, such as cryptocurrencies, and enabled more efficient and secure peer-to-peer transactions. Furthermore, the use of smart contracts has automated various processes, leading to reduced transaction times and costs. This has opened up new avenues for businesses and individuals to conduct transactions without the need for traditional intermediaries, such as banks or payment processors.

Several factors facilitated the transition of blockchain technology from hype to reality. First, the increasing number of successful use cases and pilot projects demonstrated the technology's practicality and potential for widespread adoption. As more organizations began to experiment with blockchain and DLT, the real-world applications of these technologies became more evident, encouraging further investment and development. Second, technological advancements addressed scalability and energy consumption issues, making the technology more sustainable and efficient. Innovations in consensus algorithms and network architectures have allowed for greater transaction throughput and reduced energy requirements, making blockchain and DLT more suitable for large-scale, real-world applications \cite{16}. Lastly, the growing interest and investment from both public and private sectors fueled the development and adoption of blockchain and DLT solutions. Increased awareness of the potential benefits, as well as strategic investments from major industry players, have accelerated the progress of these technologies.

Business initiatives driving the growth of DLT include consortiums and collaborations among industry players, start-ups focusing on niche applications, and the implementation of DLT in government projects. These initiatives have contributed to the adoption of DLT across various sectors, enabling organizations to reap the benefits of improved efficiency, security, and transparency \cite{m7}. Consortiums, such as R3 and the Enterprise Ethereum Alliance, bring together industry leaders to develop standards and best practices for implementing DLT solutions. By fostering collaboration and knowledge sharing, these consortiums help to accelerate the development and deployment of DLT in various industries. Meanwhile, start-ups targeting specific industry verticals have emerged, offering tailored DLT solutions to address unique challenges and opportunities. These niche players contribute to the overall growth of the DLT ecosystem by demonstrating the versatility and adaptability of the technology \cite{m6}.

In the decentralized digital age, the impact of blockchain and DLT on transactions has been profound. The traditional transaction landscape, characterized by centralized intermediaries and time-consuming processes, is being replaced by a more agile, secure, and efficient system, enabled by blockchain and DLT. As a result, the scale of transactions has grown significantly, with millions of transactions taking place daily on various blockchain networks. Furthermore, the nature of these transactions has transformed, with greater automation, programmability, and trust among participants. This transformation has been facilitated by several factors, including the growing ubiquity of digital assets, such as cryptocurrencies, and the adoption of smart contracts. The rise of digital assets has disrupted traditional financial markets and introduced new forms of value exchange, enabling individuals and organizations to conduct transactions seamlessly across borders. Additionally, the use of smart contracts has automated complex processes, reducing transaction times and costs, while also increasing trust among participants. As a result, transactions in the decentralized digital age have become more efficient, secure, and transparent than ever before \cite{m8}, \cite{m9}.

To summarize, blockchain technology and Distributed Ledger Technology have brought about a significant transformation in the way transactions are conducted in the digital age. By offering increased security, transparency, and efficiency, these technologies have disrupted traditional business models and paved the way for innovative solutions. Addressing the key questions surrounding their applications, methodologies, advantages, and disadvantages helps to build a thorough understanding of the potential impact of blockchain and DLT on the business world. As the world continues to embrace the potential of these technologies, understanding their nuances and implications will be crucial for navigating the future of transactions and the broader digital economy.

In conclusion, the widespread adoption of blockchain and Distributed Ledger Technology has transformed the transaction landscape, providing improved efficiency, security, and transparency for various industries. The transition of blockchain technology from hype to reality has been facilitated by a growing recognition of its practical applications, technological advancements addressing limitations, and increased interest and investment from both public and private sectors. As the world continues to embrace the potential of these technologies, understanding their applications, methodologies, advantages, and disadvantages will be essential for harnessing their full potential and navigating the future of the digital economy. With the ongoing development and deployment of blockchain and DLT solutions, we can expect to see even more significant changes in the way transactions are conducted and the broader impact on the global economy.

As blockchain and DLT continue to evolve, it is essential to monitor emerging trends and developments in the field. One such trend is the growing interest in interoperability between different blockchain networks and DLT systems. This would enable seamless communication and data exchange between different platforms, opening up new possibilities for collaboration and innovation. Efforts such as the Interledger Protocol (ILP) and the Polkadot network are examples of projects focused on achieving cross-chain interoperability \cite{m10}.

Another area of interest is the development of decentralized finance (DeFi) solutions. DeFi platforms leverage blockchain technology and smart contracts to provide financial services, such as lending, borrowing, and trading, without the need for traditional intermediaries. The growth of DeFi has the potential to democratize access to financial services and create new business models that challenge the dominance of traditional financial institutions. As more people gain access to these services, it is likely that the global economy will see a shift in power dynamics and a more inclusive financial landscape. In addition to financial applications, blockchain and DLT are also being explored in the fields of digital identity and privacy. Decentralized identity solutions built on blockchain technology can provide secure, verifiable, and user-controlled digital identities, potentially replacing traditional identity management systems. Such solutions could empower individuals with greater control over their personal information, reducing the risk of identity theft and improving overall privacy.

Moreover, the ongoing research into advanced cryptographic techniques, such as zero-knowledge proofs and secure multi-party computation, has the potential to further enhance the privacy and security capabilities of blockchain and DLT. Implementing these advanced techniques could lead to the development of new applications that require strong privacy guarantees, such as secure voting systems and confidential data sharing platforms. As blockchain and DLT continue to gain traction in various industries, it is also essential to consider the regulatory landscape and its impact on the growth of these technologies. Regulators around the world are grappling with the challenges posed by decentralized technologies and digital assets, seeking to strike a balance between fostering innovation and ensuring consumer protection, financial stability, and compliance with existing laws. As the regulatory environment evolves, it will be crucial for businesses and developers to stay informed and adapt their solutions accordingly \cite{m11}.

In conclusion, the transformative potential of blockchain technology and Distributed Ledger Technology is far-reaching and extends beyond the realm of transactions. As the world continues to embrace the potential of these technologies, understanding their applications, methodologies, advantages, and disadvantages will be essential for harnessing their full potential and navigating the future of the digital economy. The ongoing development and deployment of blockchain and DLT solutions will undoubtedly bring about significant changes in the way transactions are conducted, as well as broader impacts on various aspects of our lives, from finance and supply chain management to digital identity and privacy. As we continue to explore the possibilities offered by these technologies, it is crucial to maintain a comprehensive understanding of their implications and strive to develop innovative solutions that address the challenges of the decentralized digital age.

\section{blockchain technology and Distributed Ledger Technology (DLT) challenges}

Blockchain technology and Distributed Ledger Technology (DLT) have emerged as transformative forces with the potential to revolutionize a wide array of industries by offering decentralized, transparent, and secure solutions. However, as with any disruptive technology, these innovations come with their own set of challenges that must be addressed in order to fully harness their capabilities. This section aims to provide a comprehensive understanding of the challenges and potential benefits of blockchain and DLT by examining the risks associated with them, identifying suitable business applications, understanding the distributed ledger taxonomy, and exploring how DLT can enhance security and privacy methods. The distribution ledger comparison is illustrated in Figure 2.

\begin{figure*}[!ht]
\centering
\includegraphics[width=15cm,height=13cm,keepaspectratio]{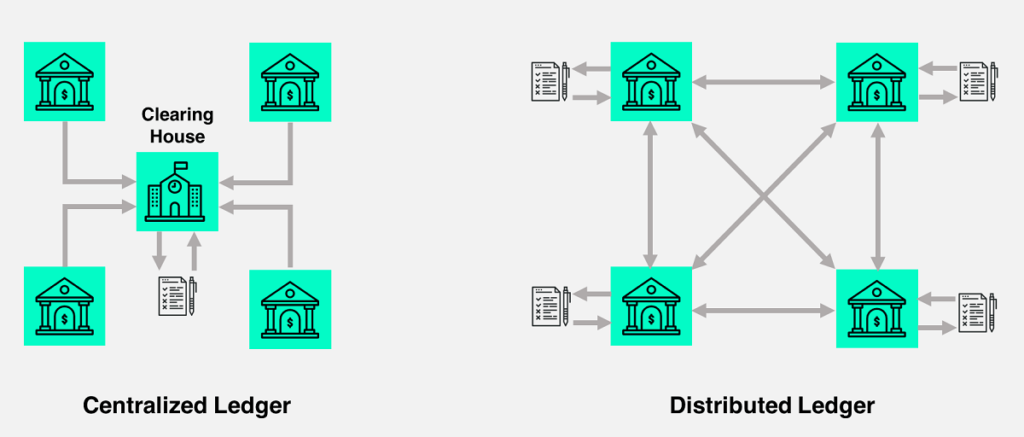}
\caption{Illustration of Ledger technologies.}
\label{fog}
\end{figure*}

\subsection{Challenges and Risks Associated with Blockchain Technology}

Blockchain technology has been hailed as a game-changer for various industries. However, it is not without its challenges, many of which are still being addressed by researchers and developers. These challenges can be broadly classified into categories such as scalability, energy consumption, regulatory compliance, interoperability, security, and privacy.

\subsubsection{Scalability}

One of the most significant challenges faced by blockchain technology is scalability. As the number of users and transactions increases, the system can become bogged down, leading to slow transaction throughput and increasing transaction costs. For example, during periods of high demand, Bitcoin and Ethereum networks have experienced congestion, resulting in delayed transactions and increased fees \cite{16}.

\subsubsection{Energy Consumption}

Blockchain networks, particularly those using Proof-of-Work (PoW) consensus algorithms, consume vast amounts of energy. This has raised sustainability concerns, as the environmental impact of mining cryptocurrencies like Bitcoin is significant. Alternative consensus mechanisms, such as Proof-of-Stake (PoS) and Delegated Proof-of-Stake (DPoS), have been proposed to address this issue.

\subsubsection{Regulatory Compliance}

The decentralized nature of blockchain technology presents challenges for regulators worldwide in ensuring compliance with existing laws, such as anti-money laundering (AML) and know-your-customer (KYC) regulations. For example, governments are grappling with how to regulate decentralized exchanges and initial coin offerings (ICOs) while still fostering innovation \cite{17}.

\subsubsection{Interoperability}

Interoperability between multiple blockchain networks and DLT systems remains an ongoing challenge. As the number of blockchain platforms increases, the need for seamless communication and interaction between these networks becomes more critical. Solutions such as cross-chain protocols and atomic swaps have been proposed to address this issue.

\subsubsection{Security}

Despite the inherent security provided by the decentralized and cryptographic nature of blockchain technology, vulnerabilities and attacks can still occur. Ensuring the robustness and resilience of blockchain networks against potential threats is essential. For instance, the infamous DAO hack in 2016, where hackers exploited a vulnerability in the smart contract code, resulted in the loss of millions of dollars' worth of Ether \cite{18}.

\subsubsection{Privacy}

Privacy concerns arise from the transparent and easily traceable transaction data on public blockchains. Developing advanced privacy-preserving techniques for blockchain technology is vital for maintaining user trust and privacy. Technologies such as zero-knowledge proofs and confidential transactions have been proposed to enhance privacy on blockchain networks.

\subsection{Areas with Good Business Fit for Blockchain Technology}

Blockchain technology has the potential to revolutionize various sectors by providing innovative solutions to long-standing problems. Some areas with a potentially good business fit for blockchain technology include financial services, supply chain management, healthcare, voting systems, intellectual property management, and decentralized identity solutions.

In financial services, blockchain can enable faster, more efficient, and secure transactions, with applications such as cross-border payments, trade finance, asset tokenization, and decentralized finance (DeFi) platforms. Blockchain can significantly improve transparency, traceability, and efficiency in supply chain management by providing an immutable, shared record of all transactions and product movements across the entire chain \cite{19}.

The healthcaresector can also benefit from blockchain technology by enhancing data security, interoperability, and patient privacy through secure, tamper-proof records of patient data and enabling data sharing among authorized parties. Blockchain technology can be utilized to develop secure, transparent, and auditable voting systems, ensuring that the voting process is free from tampering and manipulation.

Furthermore, blockchain can be used to create immutable records of intellectual property ownership, enabling secure and transparent management of copyrights, patents, and trademarks. Decentralized identity solutions built on blockchain technology can provide secure, verifiable, and user-controlled digital identities, empowering individuals with greater control over their personal information and reducing the risk of identity theft \cite{m12}.

\subsection{Distributed Ledger Taxonomy and Relation to Blockchain Technology}

Distributed ledger technology (DLT) is a digital system that facilitates the secure and transparent recording of transactions and their associated data. It enables multiple participants to maintain a shared and synchronized copy of the records, ensuring accuracy and preventing fraud. The most well-known and widely implemented form of DLT is blockchain technology. In this section, we will discuss the taxonomy of distributed ledger technologies and illustrate their relation to blockchain with real-time examples. Understanding the taxonomy of distributed ledger technology is crucial for grasping the various types of DLT systems and their relationships with blockchain technology. The distributed ledger taxonomy can be broadly classified into four categories: public distributed ledgers, private distributed ledgers, permissioned distributed ledgers, and federated or consortium ledgers. Figure is showing the texonomy of DLT \cite{20}.

\begin{figure*}[!ht]
\centering
\includegraphics[width=15cm,height=13cm,keepaspectratio]{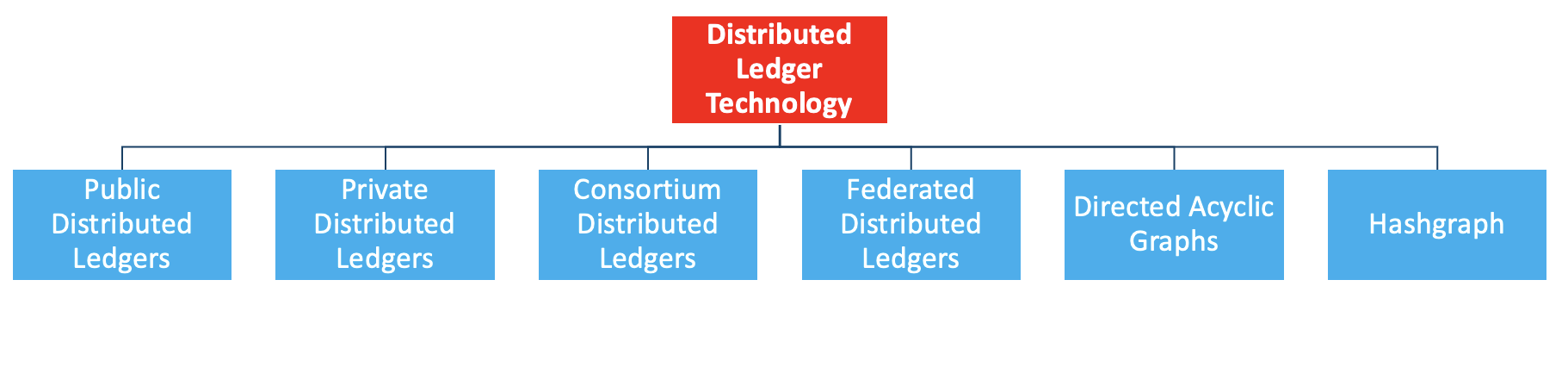}
\caption{Illustration of Taxonomy of DLT}
\label{fog}
\end{figure*}

\subsubsection{Public Distributed Ledgers}
Public distributed ledgers, also known as permissionless ledgers, are open to anyone who wishes to participate in the network. Participants can join and leave the network without seeking permission from a central authority, and they can engage in activities such as transaction validation, asset creation, or smart contract execution. Public distributed ledgers rely on consensus algorithms, such as Proof of Work (PoW) or Proof of Stake (PoS), to maintain the integrity and security of the network.

Bitcoin is the most famous example of a public distributed ledger. It employs the PoW consensus algorithm, where miners compete to solve complex mathematical problems, and the first one to solve it adds the new block of transactions to the blockchain. The Bitcoin network is open to anyone, and its transparent nature allows all participants to access and verify transactions.

\subsubsection{Private Distributed Ledgers}
Private distributed ledgers, or permissioned ledgers, require participants to obtain permission from a central authority or a consortium of entities to join the network. These ledgers offer greater control over data privacy and user access, making them suitable for organizations and industries that require strict data protection and confidentiality. Private distributed ledgers often utilize consensus algorithms such as Practical Byzantine Fault Tolerance (PBFT) or Raft, which provide faster transaction times and scalability compared to public distributed ledgers \cite{21}.

Corda, developed by R3, is a private distributed ledger platform designed for use in financial services and other regulated industries. It enables organizations to build applications that facilitate secure and private transactions. Corda's network only allows authorized participants, ensuring data confidentiality and compliance with regulatory requirements.

\subsubsection{Consortium Distributed Ledgers}
Consortium distributed ledgers are a hybrid of public and private ledgers. In these systems, multiple organizations form a consortium to govern the network, controlling access and permissions. Consortium distributed ledgers offer a balance between the transparency of public ledgers and the privacy and control of private ledgers, making them suitable for industries that require collaboration between multiple stakeholders while maintaining data privacy \cite{23}.

Quorum, a permissioned version of Ethereum, is an example of a consortium distributed ledger. It was initially developed by J.P. Morgan for use in the financial industry and is designed for use cases requiring high throughput, data privacy, and collaboration between multiple organizations.

\subsubsection{Federated Distributed Ledgers}
Federated distributed ledgers are a subtype of consortium ledgers where a group of trusted nodes, called validators, is responsible for validating and adding transactions to the ledger. This approach offers enhanced security, as validators are vetted and trusted by the network participants, and it provides faster transaction times and greater scalability compared to other DLT types.

Ripple (XRP) is a federated distributed ledger that enables fast and cost-effective cross-border payments. Ripple's network includes a set of trusted validators, which are responsible for maintaining the integrity and security of the ledger. This system provides a secure and efficient solution for international money transfers and settlements.

\subsubsection{Directed Acyclic Graphs (DAG)}
Directed Acyclic Graphs (DAG) are a type of DLT that deviates from the traditional blockchain structure. Instead of using a linear, sequential arrangement of blocks, DAGs employ a graph-like structure where transactions are interconnected, forming a directed and acyclic network. DAG-based DLTs can offer higher scalability and reduced transaction times compared to traditional blockchains, as transactions can be processed concurrently rather than sequentially.

IOTA is a prominent example of a DAG-based distributed ledger. Its Tangle network was designed to facilitate secure and feeless transactions for the Internet of Things (IoT) devices. In IOTA's Tangle, transactions are validated by the participants themselves, which removes the need for dedicated miners and reduces transaction times and costs.

\subsubsection{Hashgraph}
Hashgraph is another alternative to traditional blockchain technology. It uses a consensus algorithm called the Swirlds Consensus Algorithm, which is based on the concept of a gossip protocol. In hashgraph, transactions are shared between nodes using a "gossip about gossip" approach, where each node shares the information it has received from others, as well as information about the source of that information. This process continues until all nodes are aware of the transactions, and a consensus is reached through a virtual voting mechanism.

Hedera Hashgraph is a public distributed ledger that employs the hashgraph consensus algorithm. It aims to provide higher transaction throughput, lower latency, and increased security compared to traditional blockchain systems. Hedera Hashgraph is suited for applications requiring fast and secure transactions, such as micropayments, smart contracts, and supply chain management.

While blockchain technology is a type of distributed ledger, not all distributed ledgers are blockchains. As seen in the taxonomy above, several alternative DLTs, such as DAGs and hashgraphs, deviate from the traditional blockchain structure. However, these technologies still share some common features with blockchain, such as decentralization, immutability, and transparency. Blockchain technology has undoubtedly played a critical role in popularizing the concept of distributed ledgers, but the taxonomy of DLTs extends beyond blockchains. As the distributed ledger landscape continues to evolve, new technologies and innovations will emerge, addressing the limitations of current DLTs and enabling a wide range of real-time applications across various industries \cite{25}.

In conclusion, the taxonomy of distributed ledger technologies encompasses a diverse range of systems, each with its unique features, advantages, and use cases. Understanding the relationship between these different DLTs and blockchain technology is crucial for organizations and individuals looking to adopt and implement the most suitable solution for their specific needs. As the technology matures and advances, it will undoubtedly continue to reshape industries, drive innovation, and redefine the way we conduct transactions and share information.

\subsection{Enhancing Security and Privacy with Distributed Ledger Technology}

Distributed ledger technology (DLT) has been gaining momentum in recent years, primarily due to its potential to revolutionize various industries by enhancing security and privacy in data management and transaction processing. This section will discuss the ways in which DLT can improve security and privacy, along with real-life examples and potential challenges in implementing these technologies.

\subsubsection{Decentralization}
One of the most significant features of distributed ledger technology is its decentralized nature. Decentralization eliminates the need for a central authority, thereby reducing the risk of single points of failure and enhancing the overall security of the network. By distributing data across multiple nodes, DLT can prevent unauthorized access and tampering, as any attempt to alter the information would require compromising a majority of the nodes in the network. Decentralized finance (DeFi) platforms, such as Uniswap and Aave, leverage the decentralized nature of blockchain technology to provide financial services without intermediaries. This approach not only increases security but also democratizes access to financial services by reducing dependency on traditional centralized institutions \cite{26}, \cite{m13}.

\subsubsection{Immutability}
DLT ensures data immutability through cryptographic techniques, such as hashing and digital signatures. Once a transaction is recorded on a distributed ledger, it becomes virtually impossible to alter or delete it without detection. This feature makes DLT resistant to fraud, data tampering, and cyberattacks, ensuring the integrity and authenticity of the stored information. Supply chain management solutions, such as VeChain and IBM Food Trust, utilize blockchain's immutability to provide end-to-end traceability of products. This enhanced transparency helps in combating counterfeit goods, ensuring product authenticity, and improving overall supply chain efficiency.

\subsubsection{Encryption and Privacy}
Many distributed ledger technologies employ advanced cryptographic techniques to secure data and maintain privacy. Public and private key cryptography enables secure communication between parties while preserving the confidentiality of the transaction details. In addition, zero-knowledge proofs and other advanced privacy-preserving techniques can further enhance privacy by allowing parties to verify transactions without revealing sensitive information. Zcash, a privacy-focused cryptocurrency, uses zero-knowledge proofs called zk-SNARKs to validate transactions without revealing the sender, receiver, or transaction amount. This technology enables secure and private transactions while maintaining the integrity of the network \cite{27}.

\subsubsection{Consensus Mechanisms}
DLTs employ various consensus mechanisms to validate transactions and ensure network security. These mechanisms, such as Proof of Work (PoW), Proof of Stake (PoS), and Practical Byzantine Fault Tolerance (PBFT), enable distributed networks to reach agreement on the state of the ledger, even in the presence of malicious nodes. By requiring validators to invest resources, such as computational power or stake, consensus mechanisms make it prohibitively expensive for an attacker to manipulate the network. Ethereum, a popular blockchain platform, is transitioning from PoW to PoS consensus mechanism with Ethereum 2.0. This shift aims to enhance the security and energy efficiency of the network, making it more resistant to attacks and fostering a more sustainable ecosystem \cite{28}.

\subsubsection{Secure Smart Contracts}
Smart contracts are self-executing agreements with the terms directly written into code. They run on distributed ledger platforms, enabling automatic enforcement of contractual obligations without the need for intermediaries. By leveraging cryptographic techniques and consensus mechanisms, smart contracts can offer increased security and transparency, reducing the risk of fraud and disputes. The insurance industry has begun exploring the use of smart contracts to automate claims processing. Companies like Etherisc and Aigang use blockchain-based smart contracts to process claims and payouts, streamlining the process and reducing the potential for fraudulent claims.

\subsection{Challenges in Enhancing Security and Privacy with DLT}

Despite the numerous advantages that DLT offers in enhancing security and privacy, there are challenges that need to be addressed for widespread adoption and implementation of these technologies.

\subsubsection{Scalability}
Scalability remains a significant challenge for many distributed ledger technologies, particularly those utilizing blockchain. As the number of transactions and participants increases, networks can become congested, leading to slower transaction times and increased costs. Developing scalable solutions without compromising security and privacy is crucial for DLT to achieve widespread adoption and cater to the needs of various industries.

\subsubsection{Interoperability}
The rapidly growing landscape of distributed ledger technologies has resulted in numerous disparate platforms and protocols, often with limited compatibility. Interoperability between different DLTs is essential for seamless data exchange and collaboration across various systems and industries. Standardization and development of cross-chain solutions are crucial to ensure that security and privacy enhancements provided by DLT can be fully leveraged across different networks \cite{28}, \cite{m14}.

\subsubsection{Regulatory and Legal Frameworks}
As distributed ledger technology continues to evolve and find applications across various sectors, the need for clear regulatory and legal frameworks becomes increasingly important. Policymakers need to strike a balance between fostering innovation and ensuring that DLT-based solutions comply with existing laws and regulations, particularly concerning data privacy and security. Harmonizing the regulatory landscape will be essential for building trust and encouraging the adoption of DLT.

\subsubsection{Education and Awareness}
The adoption of distributed ledger technology requires a significant shift in mindset for many organizations and individuals. Educating stakeholders about the benefits and potential risks associated with DLT is essential to address misconceptions and foster informed decision-making. Building awareness and promoting collaboration between developers, users, and regulators can help drive the adoption of secure and privacy-enhancing DLT solutions \cite{m15}.

\subsubsection{Technological Advancements}
The distributed ledger technology landscape is continuously evolving, with new innovations emerging to address the existing limitations and enhance security and privacy features. It is crucial for organizations and developers to stay updated with the latest advancements, invest in research and development, and be prepared to adapt to the changing technological landscape.

In conclusion, distributed ledger technology has the potential to significantly enhance security and privacy across various industries and applications. By addressing the challenges associated with scalability, interoperability, regulatory frameworks, education and awareness, and technological advancements, DLT can revolutionize the way we manage and secure data, enabling a more trustworthy and efficient digital ecosystem. As organizations and individuals continue to adopt and implement DLT-based solutions, it is essential to prioritize security and privacy to ensure the long-term success and sustainability of this transformative technology \cite{29}. Blockchain technology and Distributed Ledger Technology (DLT) have the potential to transform various industries by offering decentralized, transparent, and secure solutions. However, realizing the full potential of these technologies requires addressing the challenges and risks associated with them. By examining the current challenges, identifying suitable business applications, understanding the distributed ledger taxonomy, and exploring the ways in which DLT can enhance security and privacy methods, we can gain a comprehensive understanding of the potential benefits and limitations of blockchain and DLT. This understanding will be essential in guiding future research, development, and implementation of these groundbreaking technologies \cite{30}.

\section{Role of Distributed Ledger Technology (DLT) and blockchain in intelligence sharing}

Intelligence sharing is a crucial component in ensuring the security of individuals, organizations, and nations. However, this process often faces challenges such as a lack of transparency, trust, and interoperability. Distributed Ledger Technology (DLT) and blockchain have emerged as potential solutions to address these issues. These technologies offer transparency, trust, immutability, and decentralization, making them well-suited for enhancing intelligence sharing. In this paper, we will explore the impact of DLT and blockchain on intelligence sharing. Figure is showing the architecture of intelligence sharing in blockchain.

\begin{figure*}[!ht]
\centering
\includegraphics[width=15cm,height=13cm,keepaspectratio]{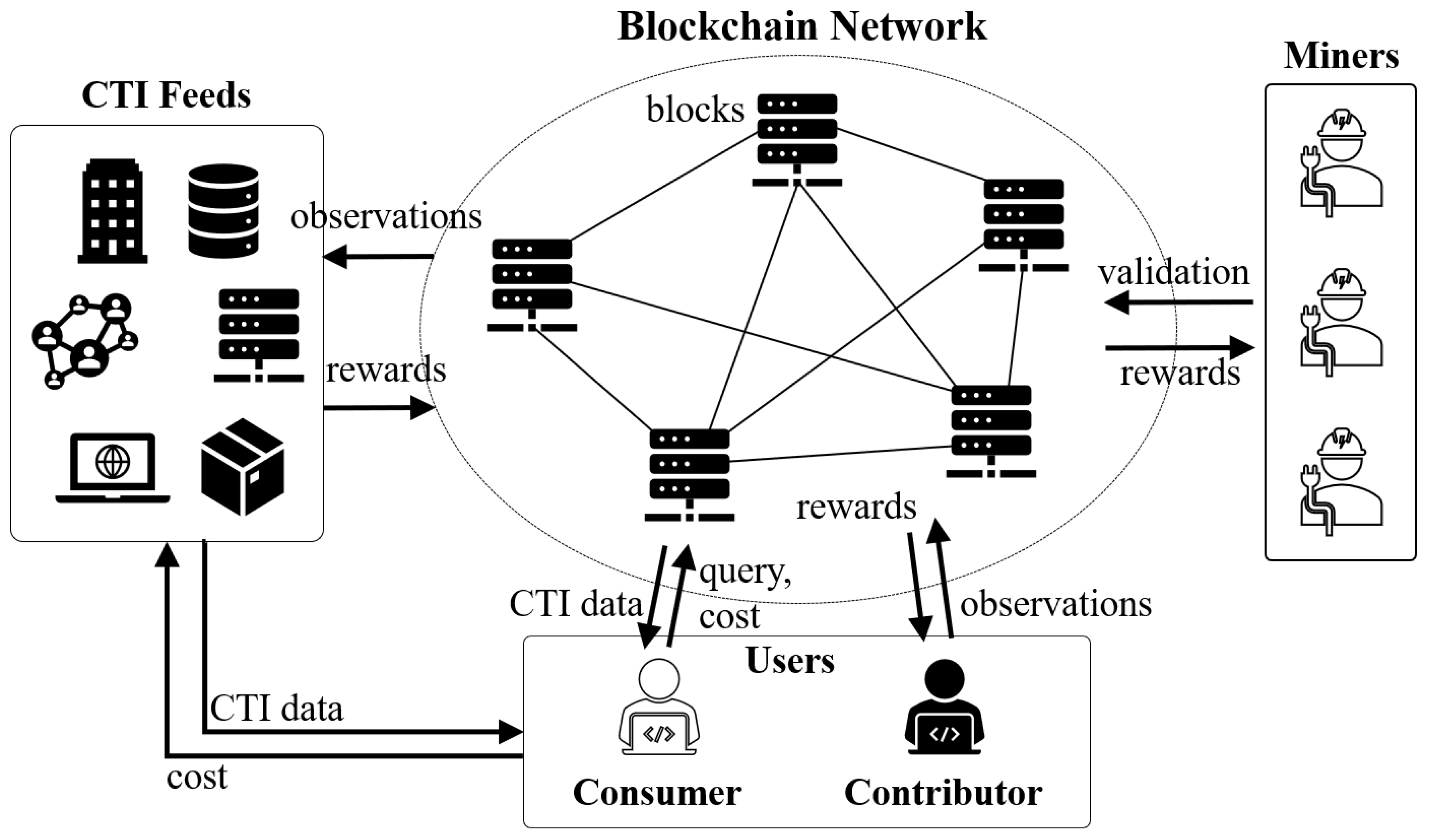}
\caption{Illustration intelligence sharing using blockchain}
\label{fog}
\end{figure*}

\subsection{Enhancing Intelligence Sharing with DLT and Blockchain}

DLT and blockchain can enhance intelligence sharing by leveraging their inherent features of decentralization, transparency, immutability, and security. The following examples demonstrate the potential of these technologies in intelligence sharing.

\subsubsection{Secure Communication Platforms} DLT and blockchain can create secure communication platforms to ensure that sensitive information is exchanged only between authorized parties. These platforms provide confidentiality, integrity, and non-repudiation, mitigating the risks of data breaches and unauthorized access. For example, the European Union has implemented the blockchain-based platform called EU Blockchain Initiative to enhance secure communication among member states. This platform enables secure data exchange and helps prevent cyber threats.

\subsubsection{Cyber Threat Intelligence Sharing} DLT and blockchain facilitate the sharing of cyber threat intelligence among organizations, enabling them to collaborate and respond more effectively to cyber attacks. These technologies ensure that threat information is securely and efficiently disseminated across the network, promoting real-time situational awareness and enabling proactive defense measures. For example, the Cyber Threat Intelligence Network (CTIN) leverages DLT to provide a secure and decentralized environment for threat intelligence sharing.

\subsubsection{Identity Management} DLT and blockchain can help establish a trusted and decentralized identity management system that enables intelligence sharing. These technologies can provide secure and tamper-proof storage of identity information, ensuring privacy and data protection. For example, the Self Sovereign Identity (SSI) initiative leverages blockchain to provide a decentralized identity management system that empowers individuals to control their identity data and share it securely.

\subsubsection{Supply Chain Security} DLT and blockchain can track and authenticate goods in global supply chains, ensuring the integrity and security of products. These technologies can provide end-to-end visibility and a tamper-proof record of product movement, combatting counterfeit products and detecting potential security threats. For example, IBM has implemented a blockchain-based supply chain management system that enables secure and transparent tracking of goods across the supply chain.

\subsection{Potential Benefits and Challenges}

The implementation of DLT and blockchain in intelligence sharing offers several potential benefits:

\subsubsection{Enhanced Security} DLT and blockchain provide improved security through decentralization, cryptography, and consensus mechanisms, ensuring data integrity and protecting against tampering and unauthorized access.

\subsubsection{Increased Efficiency} By eliminating intermediaries and automating processes through smart contracts, DLT and blockchain reduce transaction costs and enhance efficiency in intelligence sharing.

\subsubsection{Transparency and Accountability} DLT and blockchain provide transparency and accountability by creating a tamper-proof record of transactions that can be audited and verified by all network participants.

However, there are several challenges associated with implementing DLT and blockchain in intelligence sharing:

\subsubsection{Interoperability} Different DLT and blockchain systems may have different standards and protocols, making it difficult to integrate them with existing systems.

\subsubsection{Scalability} The current infrastructure of DLT and blockchain may not be able to handle the volume of transactions required for intelligence sharing, leading to slow processing times and high fees.

\subsubsection{Regulation} The lack of regulation and standards in the use of DLT and blockchain in intelligence sharing may result in legal and compliance issues.

\subsubsection{Privacy Concerns} The inherent transparency of DLT and blockchain may pose privacy concerns in intelligence sharing, as sensitive information may be accessible to all network participants.

\subsection{Impact on Privacy and Potential Risks}

While DLT and blockchain offer significant benefits in intelligence sharing, they also have a considerable impact on privacy and carry potential risks.

\subsubsection{Impact on Privacy} DLT and blockchain can impact privacy in intelligence sharing by creating a tamper-proof record of transactions that is accessible to all network participants. While this enhances transparency and accountability, it also poses privacy concerns, as sensitive information may be disclosed. For example, in a blockchain-based supply chain management system, all participants in the supply chain can view the details of each transaction, including the products and their origins. While this enhances transparency, it may also reveal sensitive information about the parties involved, such as their business relationships and processes.

\subsubsection{Potential Risks} DLT and blockchain in intelligence sharing carry potential risks, such as cyber threats, data breaches, and attacks on the consensus mechanisms. For example, a cyber attack on a blockchain-based communication platform may compromise the confidentiality and integrity of the data shared through the platform. Similarly, a data breach in a blockchain-based identity management system may result in the unauthorized access and use of personal information \cite{m16}.

\subsubsection{Mitigating Risks} To mitigate the risks associated with DLT and blockchain in intelligence sharing, organizations must implement appropriate security measures, such as strong authentication mechanisms, data encryption, and multi-factor authentication. Additionally, regulatory frameworks and standards must be developed to address the legal and compliance issues associated with the use of these technologies. For example, the General Data Protection Regulation (GDPR) in the European Union provides a legal framework for the protection of personal data, including data stored on blockchains.

In conclusion, DLT and blockchain offer significant potential in enhancing intelligence sharing, providing secure communication platforms, cyber threat intelligence sharing, identity management, and supply chain security. However, the challenges of interoperability, scalability, regulation, and privacy must be addressed to enable widespread adoption. The impact on privacy and potential risks associated with DLT and blockchain must also be taken into account, and appropriate security measures must be implemented to mitigate these risks. Overall, the use of DLT and blockchain in intelligence sharing requires a careful balance between innovation and risk management.
\begin{table*}[]
\setlength\tabcolsep{0pt}
\caption{Traditional methods of Intelligence sharing}
\begin{tabular}{|c|c|c|c|}
\hline
\textbf{Method}                                                                               & \textbf{Description}                                                                                                                                     & \textbf{Advantages}                                                                                                                         & \textbf{Disadvantages}                                                                                                           \\ \hline
Human Intelligence (HUMINT)                                                                   & \begin{tabular}[c]{@{}c@{}}Collecting information directly \\ from human sources\end{tabular}                                                            & \begin{tabular}[c]{@{}c@{}}Provides first-hand information; \\ Access to information from \\ individuals with direct knowledge\end{tabular} & \begin{tabular}[c]{@{}c@{}}Subject to human error; \\ Requires establishing trust \\ and credibility with sources\end{tabular}   \\ \hline
Signals Intelligence (SIGINT)                                                                 & \begin{tabular}[c]{@{}c@{}}Interception and analysis of \\ electronic communications\end{tabular}                                                        & \begin{tabular}[c]{@{}c@{}}Provides valuable information \\ without direct contact; \\ Can monitor target communications\end{tabular}       & \begin{tabular}[c]{@{}c@{}}Requires sophisticated equipment; \\ Encrypted communications \\ can be unreadable\end{tabular}       \\ \hline
Imagery Intelligence (IMINT)                                                                  & \begin{tabular}[c]{@{}c@{}}Analysis of visual images \\ (satellite imagery, aerial photographs)\end{tabular}                                             & \begin{tabular}[c]{@{}c@{}}Valuable information on activities \\ without direct contact; \\ Can monitor target activities\end{tabular}      & \begin{tabular}[c]{@{}c@{}}Requires sophisticated equipment; \\ Images can be altered or \\ manipulated\end{tabular}             \\ \hline
\begin{tabular}[c]{@{}c@{}}Measurement and Signature Intelligence \\ (MASINT)\end{tabular}    & \begin{tabular}[c]{@{}c@{}}Collection and analysis of data\\  from non-standard sources\\  (radar, sensors)\end{tabular}                                 & \begin{tabular}[c]{@{}c@{}}Provides technical intelligence \\ on potential threats\end{tabular}                                             & \begin{tabular}[c]{@{}c@{}}Specialized method; \\ Requires advanced equipment and \\ expertise\end{tabular}                      \\ \hline
Open-Source Intelligence (OSINT)                                                              & \begin{tabular}[c]{@{}c@{}}Collection and analysis of publicly \\ available sources \\ (news articles, social media, \\ government reports)\end{tabular} & \begin{tabular}[c]{@{}c@{}}Access to a wealth of information \\ without direct contact\end{tabular}                                         & \begin{tabular}[c]{@{}c@{}}Requires verification of accuracy \\ and credibility\end{tabular}                                     \\ \hline
Diplomatic Channels                                                                           & \begin{tabular}[c]{@{}c@{}}Intelligence sharing between \\ governments through embassies\end{tabular}                                                    & \begin{tabular}[c]{@{}c@{}}Confidential communication \\ between nations\end{tabular}                                                       & \begin{tabular}[c]{@{}c@{}}Trust and confidentiality issues; \\ Limited to state actors\end{tabular}                             \\ \hline
Military-to-Military Exchanges                                                                & \begin{tabular}[c]{@{}c@{}}Sharing military intelligence \\ between allied nations\end{tabular}                                                          & \begin{tabular}[c]{@{}c@{}}Effective for sharing intelligence \\ related to military operations\end{tabular}                                & \begin{tabular}[c]{@{}c@{}}Limited to military threats \\ and operations\end{tabular}                                            \\ \hline
Law Enforcement Agencies                                                                      & \begin{tabular}[c]{@{}c@{}}Intelligence sharing through \\ agencies like the \\ FBI, DHS, NSA\end{tabular}                                               & \begin{tabular}[c]{@{}c@{}}Coordinated efforts in addressing \\ security threats\end{tabular}                                               & \begin{tabular}[c]{@{}c@{}}Potential gaps in communication; \\ Inter-agency rivalry\end{tabular}                                 \\ \hline
\begin{tabular}[c]{@{}c@{}}Multinational Organizations \\ (e.g., Interpol, NATO)\end{tabular} & \begin{tabular}[c]{@{}c@{}}Platform for member countries \\ to share intelligence and \\ coordinate efforts\end{tabular}                                 & \begin{tabular}[c]{@{}c@{}}Streamlined communication and \\ coordination; \\ Addresses global security threats\end{tabular}                 & \begin{tabular}[c]{@{}c@{}}Requires trust between member nations; \\ Sharing limitations due to national interests\end{tabular}  \\ \hline
Public-Private Sector Sharing                                                                 & \begin{tabular}[c]{@{}c@{}}Sharing intelligence between \\ government agencies and \\ private sector companies\end{tabular}                              & \begin{tabular}[c]{@{}c@{}}Access to valuable intelligence \\ on cyber threats and other attacks\end{tabular}                               & \begin{tabular}[c]{@{}c@{}}Trust and confidentiality issues; \\ Varying standards and systems between organizations\end{tabular} \\ \hline
\end{tabular}
\end{table*}

\section{Intelligence sharing}

\subsection{Traditional methods used for intelligence sharing}

Intelligence sharing is a crucial component of national security, as it enables different organizations to exchange information and collaborate in the fight against potential threats and criminal activities. Traditional methods of intelligence sharing have been used for many years, and they have undergone significant refinement and evolution over time. In this section, we will discuss some of the most commonly used traditional methods of intelligence sharing and the challenges associated with them. Traditional methods of intelligence sharing is described in table 1.

Human Intelligence (HUMINT) \cite{31} is one of the most traditional methods of intelligence sharing. It involves collecting information directly from human sources. The advantage of HUMINT is that it can provide first-hand information from individuals who have direct access to the information being sought. However, HUMINT is also subject to various limitations, such as the potential for human error and the need to establish trust and credibility with sources.

Signals Intelligence (SIGINT) \cite{32} involves the interception and analysis of electronic communications, such as radio signals and emails. This method is often used by intelligence agencies to monitor the communications of potential threats and gain insights into their activities. The primary advantage of SIGINT is that it can provide valuable information without the need for direct contact with the target. However, SIGINT is also subject to various limitations, such as the need for sophisticated equipment and the potential for encryption to render the intercepted communications unreadable.

Imagery Intelligence (IMINT) \cite{33} involves the analysis of visual images, such as satellite imagery and aerial photographs. This method is often used by intelligence agencies to monitor the activities of potential threats and gain insights into their capabilities. The primary advantage of IMINT is that it can provide valuable information on the activities of potential threats without the need for direct contact. However, IMINT is also subject to various limitations, such as the need for sophisticated equipment and the potential for images to be altered or manipulated.

Measurement and Signature Intelligence (MASINT) \cite{34} is a specialized form of intelligence that involves the collection and analysis of data from non-standard sources, such as radar and other types of sensors. This method is often used to provide intelligence on technical aspects of potential threats, such as their weapons systems or communication networks.

Open-Source Intelligence (OSINT) \cite{35} involves the collection and analysis of information from publicly available sources, such as news articles, social media, and government reports. The primary advantage of OSINT is that it can provide a wealth of information without the need for direct contact with the target. However, OSINT is also subject to various limitations, such as the need to verify the accuracy and credibility of the information collected.

Collaboration and information sharing between various intelligence agencies and other organizations are also essential components of intelligence sharing. This involves the sharing of information and resources to enable a comprehensive understanding of potential threats and criminal activities.

One significant challenge associated with traditional methods of intelligence sharing is the need to manage and integrate information from various sources, including those from different organizations and countries. This requires a significant investment in information technology infrastructure, including secure communication networks and sophisticated data analysis tools. Another challenge is the potential for information overload, where too much data can lead to difficulties in identifying and prioritizing potential threats.

Moreover, traditional methods of intelligence sharing are often reactive, meaning that they are designed to respond to threats after they have emerged. This approach may not be sufficient in a rapidly changing security landscape, where threats can emerge and evolve quickly. Therefore, intelligence agencies are increasingly turning to new methods and technologies, such as big data analytics and artificial intelligence (AI), to enable a more proactive approach to intelligence gathering and sharing.

In summary, traditional methods of intelligence sharing have been used for many years and have undergone significant refinement and evolution. These methods provide valuable insights into potential threats and criminal activities. However, they also face various challenges, such as the need to manage and integrate information from multiple sources and the potential for information overload. Intelligence agencies must continue to The sharing of intelligence is essential for the prevention and detection of various forms of threats to security, ranging from cyber attacks to terrorism. In the past, intelligence sharing relied on traditional methods such as face-to-face meetings, phone calls, and written reports. However, with the rise of technology, intelligence sharing has evolved and become more complex.

One of the traditional methods of intelligence sharing is through diplomatic channels, where intelligence is shared between governments through their respective embassies. This method is still widely used today, as it allows for confidential communication between nations. Another traditional method is through military-to-military exchanges, where military intelligence is shared between allied nations. This method is effective for sharing intelligence related to military operations, but may not be suitable for other types of threats.

Intelligence sharing can also be facilitated through law enforcement agencies. In the United States, the Federal Bureau of Investigation (FBI) serves as the lead agency for counterterrorism intelligence sharing. Other law enforcement agencies, such as the Department of Homeland Security and the National Security Agency, also play a role in intelligence sharing.

In addition to these traditional methods, intelligence sharing can also occur through multinational organizations such as Interpol or the North Atlantic Treaty Organization (NATO). These organizations provide a platform for member countries to share intelligence and coordinate their efforts to address security threats.

In recent years, there has been a growing trend towards information sharing between public and private sectors. Private sector companies can provide valuable intelligence related to cyber threats and other types of attacks. The sharing of this information can help government agencies better understand the nature of the threat and develop more effective countermeasures.

However, there are also challenges associated with traditional methods of intelligence sharing. One major challenge is the issue of trust between nations. Governments may be hesitant to share intelligence with other countries due to concerns about leaks or misuse of the information. There are also concerns about the reliability and accuracy of the intelligence being shared.

Another challenge is the difficulty in sharing information across different systems and platforms. Many countries and organizations use different technologies and software for their intelligence operations, making it difficult to integrate and share information. This can result in gaps in intelligence and a lack of coordination between agencies.

In conclusion, traditional methods of intelligence sharing have played a crucial role in national security efforts for many years. However, with the evolution of technology, intelligence sharing has become more complex, and there is a growing need for innovative solutions that can facilitate the sharing of information across different platforms and systems. The challenges associated with traditional methods of intelligence sharing highlight the need for new approaches and technologies that can overcome these barriers and improve the efficiency and effectiveness of intelligence operations.

\section{Related Work}

The challenge of sharing cyber threat intelligence (CTI) lies in the potential legal and financial repercussions that organizations face, leading to limited data in terms of volume, quality, and timeliness for cybersecurity awareness and mitigation. To address this issue, the authors suggest employing a distributed blockchain ledger to enable secure sharing of CTI while allowing non-attributable participation within a threat-sharing community \cite{r1}. Drawing inspiration from Distributed Anonymous Payment (DAP) schemes in cryptocurrency, a novel token-based authentication method is introduced for use in a permissioned blockchain. This approach facilitates a consortium of semi-trusted entities to collaboratively curate CTI, ultimately benefiting the entire community.

Addressing the need for a secure and trusted framework for threat analysis and sharing, the authors propose a solution combining Hyperledger Fabric and IPFS, based on the MITRE ATT\&CK framework. Focusing on threats in Healthcare IT and other organizations, this method ensures security, privacy, and anonymity while maintaining high throughput and scalability \cite{r2}. The infrastructure employs the MITRE ATT\&CK framework, pluggable certificate authorities, and self-executing chaincode to foster trust and enhance system security. Future work includes developing a comprehensive proof-of-concept using a Kubernetes cluster in a cloud infrastructure for improved scalability.

The energy sector faces sophisticated cyberattacks, and the need for standardized, secure, and efficient cyber threat intelligence sharing is paramount. Current solutions, such as the TAXII protocol, lack adequate data integrity assurance and compatibility with event-driven architectures. The authors introduces a novel approach for secure, real-time exchange of cyber threat information by extending the TAXII framework and integrating Distributed Ledger Technologies (DLT) and a generalized publish-subscribe middleware \cite{r3}. This combination addresses data integrity and audit trail concerns and facilitates near real-time information exchange. The proposed solution's applicability is validated through multiple use cases in Electrical Power and Energy Systems, demonstrating secure, tamper-proof, and scalable cyber threat information sharing.

The increasing prevalence of DDoS attacks, particularly following the release of the Mirai botnet source code, poses significant challenges to internet-based services. Detection and mitigation of these attacks are often reactive and costly. Authors in \cite{r4} proposes a proactive, low-cost IoT botnet detection system that identifies anomalies in IoT device behavior and mitigates DDoS botnet exploitation at the source. Additionally, the study presents a collaborative trust relationship-based threat intelligence-sharing mechanism to protect other IoT devices from detected botnets. The mechanism's performance was evaluated using Ethereum Virtual Machine and Hyperledger, with a 97\% detection rate for Mirai botnet activities and greater scalability through Ethereum Virtual Machine. The research employs smart contracts and blockchain-based methodologies for trust establishment and collaboration, enabling a proactive defense against IoT botnets.

Traditional threat information sharing methods have relied on manual modeling and centralized systems, which can be inefficient and insecure. To address these issues, \cite{r5} proposes a privacy-preserving mechanism for sharing threat information using Hyperledger Fabric private-permissioned distributed ledger technology and the MITRE ATT\&CK framework. This approach enhances organizational security and automation by improving data quality, traceability, and system reliability. It also has potential applications in combating intellectual property theft and industrial espionage. The paper provides a proof-of-concept implementation, security analysis, and performance experiments to demonstrate the feasibility and effectiveness of the proposed solution. Future work includes developing a comprehensive cloud-based implementation using a Kubernetes cluster to further improve system throughput and scalability.

Authors in \cite{r6} introduces a novel trust taxonomy for creating a trusted threat sharing environment in cyber threat intelligence sharing. By analyzing and comparing 30 popular threat intelligence platforms/providers and their trust functionalities, the paper aims to enhance trust establishment and automation in sharing sensitive vulnerability information among decentralized stakeholders without compromising security.

This research work in \cite{r7} examines existing cyber threat intelligence frameworks to identify key components that form the basis for solution design. By offering a deeper understanding of these architectural designs, the study aims to assist cybersecurity practitioners in tailoring solutions to meet their organization's specific requirements.

Authors investigates the frameworks for cyber threat intelligence sharing in the United States \cite{r8}. The study evaluates the effectiveness of these frameworks, revealing potential areas for improvement to enhance collaboration and information sharing. By examining the current state of threat intelligence sharing, the paper provides valuable insights and recommendations to strengthen security measures and develop more effective strategies against cyber threats.

Authors in \cite{r9} explore the landscape of threat intelligence sharing platforms, highlighting the increased willingness of organizations to exchange information on vulnerabilities, threats, incidents, and mitigation strategies. However, the effectiveness of these platforms remains unclear due to the lack of a common definition and empirical research. The authors conducted a systematic study of 22 threat intelligence sharing platforms, comparing their features and capabilities. By identifying gaps and presenting emerging research perspectives, the study provides valuable insights for software vendors and researchers to enhance the effectiveness of these platforms and foster better information sharing practices to combat cyber threats.

In research work \cite{r10}, the authors address the challenge of trust in threat intelligence sharing by enhancing the TATIS security framework, which provides fine-grained protection for threat intelligence platform APIs. They make TATIS fully distributed, supporting federated authentication and authorization across domains, and integrate distributed ledger technology (DLT) to ensure verifiability, data provenance, and secure access control. The improved framework is implemented on the Malware Information Sharing Platform (MISP) and tested using real open-source cyber threat intelligence (CTI) data. The results demonstrate the feasibility and performance of the solution, reinforcing trustworthiness in CTI sharing through reliable access control, secure data sharing, and provenance management.

This study in \cite{r11} proposes a new blockchain network model that enables the secure dissemination of Cyber Threat Intelligence (CTI) data while addressing the trust barriers and data privacy issues inherent in the domain. Motivated by recent changes in information security legislation in the European Union and the challenges faced by Computer Security and Incident Response Teams (CSIRT) when sharing sensitive data, the authors designed a CTI sharing model using the security properties of blockchain. They implemented a testbed using Hyperledger Fabric and the STIX 2.0 protocol, successfully demonstrating the sharing of security data in a trustless environment. The prototype also achieved network partitioning and enforced sharing rules through Fabric channels and smart contracts. Future work will focus on the performance and security aspects of the CTI blockchain network for real-world applications.

The IT community faces an ongoing challenge of new threats and security incidents that are difficult to combat individually. Sharing information about these threats has become crucial for effective incident response. The authors introduces the Malware Information Sharing Platform (MISP) and threat sharing project, a trusted platform designed to collect and share indicators of compromise (IoC), vulnerabilities, and other threat information relevant to targeted attacks and fraud cases \cite{r12}. MISP aims to facilitate the establishment of preventive actions and countermeasures through collaborative knowledge sharing, ultimately enabling better detection and response to existing malware and various threats.

Cyber threat intelligence (CTI) exchange has the potential to improve societal security, but participants are often hesitant to share their CTI in voluntary-based approaches. To encourage dynamic information sharing, authors proposes a paradigm shift in cybersecurity information exchange \cite{r13}. This new approach supports the deployment of dynamic risk management frameworks and offers incentives for participants to share, invest, and consume threat intelligence and risk intelligence information. The proposal utilizes standards like Structured Threat Information Exchange and W3C semantic web standards for behavioral threat intelligence patterning. Furthermore, it introduces an Ethereum Blockchain Smart Contract Marketplace to incentivize sharing and establishes a standard CTI token as a valuable digital asset. Simulations and experimentation demonstrate the benefits, incentives, and potential limitations of this approach in terms of storage and transaction costs.

\begin{table*}[]
\setlength\tabcolsep{0pt}
\caption{Related work in Intelligence sharing using blockchain and DLT}
\begin{tabular}{|c|c|c|c|c|}
\hline
\textbf{Reference}          & \textbf{\begin{tabular}[c]{@{}c@{}}Proposed Work Methodology \\ with Algorithm Used\end{tabular}}                                                                                                 & \textbf{Problem Discussed}                                                                                                                                               & \textbf{Benefits and Outcomes}                                                                                                                     & \textbf{\begin{tabular}[c]{@{}c@{}}Application and Technology \\ Used\end{tabular}}                                                                   \\ \hline
\cite{r1}  & \begin{tabular}[c]{@{}c@{}}Distributed blockchain ledger \\ and token-based authentication\end{tabular}                                                                                           & \begin{tabular}[c]{@{}c@{}}Legal and financial \\ repercussions of sharing CTI\end{tabular}                                                                              & \begin{tabular}[c]{@{}c@{}}Secure CTI sharing; consortium of \\ semi-trusted entities; collaboration\end{tabular}                                  & \begin{tabular}[c]{@{}c@{}}Distributed blockchain \\ ledger\end{tabular}                                                                              \\ \hline
\cite{r2} & \begin{tabular}[c]{@{}c@{}}Hyperledger Fabric and IPFS combined \\ with MITRE ATT\&CK framework\end{tabular}                                                                                      & \begin{tabular}[c]{@{}c@{}}Need for secure and trusted framework \\ for threat analysis and sharing\end{tabular}                                                         & \begin{tabular}[c]{@{}c@{}}Security, privacy, anonymity, \\ high throughput, and scalability\end{tabular}                                          & \begin{tabular}[c]{@{}c@{}}Hyperledger Fabric, \\ IPFS, MITRE ATT\&CK \\ framework\end{tabular}                                                       \\ \hline
\cite{r3}  & \begin{tabular}[c]{@{}c@{}}Extended TAXII framework integrated \\ with DLT and publish-subscribe middleware\end{tabular}                                                                          & \begin{tabular}[c]{@{}c@{}}Inadequate data integrity assurance \\ and compatibility in \\ energy sector CTI sharing\end{tabular}                                         & \begin{tabular}[c]{@{}c@{}}Secure, tamper-proof, and scalable \\ CTI sharing; real-time exchange\end{tabular}                                      & \begin{tabular}[c]{@{}c@{}}TAXII framework, \\ Distributed Ledger \\ Technologies\end{tabular}                                                        \\ \hline
\cite{r4}  & \begin{tabular}[c]{@{}c@{}}IoT botnet detection system and collaborative \\ trust relationship-based sharing mechanism\end{tabular}                                                               & \begin{tabular}[c]{@{}c@{}}Detection and mitigation of \\ DDoS attacks in IoT devices\end{tabular}                                                                       & \begin{tabular}[c]{@{}c@{}}Proactive defense, 97\% detection rate\\  for Mirai botnet, scalability\end{tabular}                                    & \begin{tabular}[c]{@{}c@{}}Ethereum Virtual Machine, \\ Hyperledger\end{tabular}                                                                      \\ \hline
\cite{r5}  & \begin{tabular}[c]{@{}c@{}}Privacy-preserving mechanism using \\ Hyperledger Fabric and \\ MITRE ATT\&CK framework\end{tabular}                                                                   & \begin{tabular}[c]{@{}c@{}}Inefficient and insecure traditional \\ threat information sharing methods\end{tabular}                                                       & \begin{tabular}[c]{@{}c@{}}Improved data quality, traceability, \\ system reliability; combating IP theft \\ and industrial espionage\end{tabular} & \begin{tabular}[c]{@{}c@{}}Hyperledger Fabric, \\ MITRE ATT\&CK \\ framework\end{tabular}                                                             \\ \hline
\cite{r6}  & \begin{tabular}[c]{@{}c@{}}Novel trust taxonomy for \\ CTI sharing environment\end{tabular}                                                                                                       & \begin{tabular}[c]{@{}c@{}}Enhancing trust establishment and \\ automation in CTI sharing\end{tabular}                                                                   & \begin{tabular}[c]{@{}c@{}}Strengthening trust in sharing \\ sensitive vulnerability information \\ without compromising security\end{tabular}     & Trust taxonomy                                                                                                                                        \\ \hline
\cite{r7}  & \begin{tabular}[c]{@{}c@{}}Examination of existing \\ CTI frameworks\end{tabular}                                                                                                                 & \begin{tabular}[c]{@{}c@{}}Identifying key components for \\ solution design in CTI frameworks\end{tabular}                                                              & \begin{tabular}[c]{@{}c@{}}Assisting cybersecurity practitioners \\ in tailoring solutions\end{tabular}                                            & \begin{tabular}[c]{@{}c@{}}CTI framework \\ analysis\end{tabular}                                                                                     \\ \hline
\cite{r8}  & \begin{tabular}[c]{@{}c@{}}Investigation of US CTI \\ sharing frameworks\end{tabular}                                                                                                             & \begin{tabular}[c]{@{}c@{}}Evaluating effectiveness of US CTI \\ sharing frameworks\end{tabular}                                                                         & \begin{tabular}[c]{@{}c@{}}Strengthening security measures \\ and developing better strategies \\ against cyber threats\end{tabular}               & \begin{tabular}[c]{@{}c@{}}CTI framework \\ evaluation\end{tabular}                                                                                   \\ \hline
\cite{r9}  & \begin{tabular}[c]{@{}c@{}}Systematic study of 22 threat \\ intelligence sharing platforms\end{tabular}                                                                                           & \begin{tabular}[c]{@{}c@{}}Lack of a common definition and\\  empirical research on \\ platform effectiveness\end{tabular}                                               & \begin{tabular}[c]{@{}c@{}}Enhancing platform effectiveness \\ and fostering better \\ information sharing practices\end{tabular}                  & \begin{tabular}[c]{@{}c@{}}Threat intelligence sharing \\ platform analysis\end{tabular}                                                              \\ \hline
\cite{r10} & \begin{tabular}[c]{@{}c@{}}Enhanced TATIS security \\ framework integrated with DLT\end{tabular}                                                                                                  & \begin{tabular}[c]{@{}c@{}}Trust challenges in threat intelligence \\ sharing\end{tabular}                                                                               & \begin{tabular}[c]{@{}c@{}}Reliable access control, \\ secure data sharing, and \\ provenance management\end{tabular}                              & \begin{tabular}[c]{@{}c@{}}TATIS security framework, \\ Distributed Ledger Technology, \\ Malware Information \\ Sharing Platform (MISP)\end{tabular} \\ \hline
\cite{r11} & \begin{tabular}[c]{@{}c@{}}Blockchain network \\ model for CTI sharing\end{tabular}                                                                                                               & \begin{tabular}[c]{@{}c@{}}Trust barriers and data privacy \\ issues in CTI sharing\end{tabular}                                                                         & \begin{tabular}[c]{@{}c@{}}Secure dissemination of CTI data; \\ network partitioning; \\ sharing rules enforcement\end{tabular}                    & \begin{tabular}[c]{@{}c@{}}Hyperledger Fabric, \\ STIX 2.0 protocol\end{tabular}                                                                      \\ \hline
\cite{r12} & MISP threat sharing platform                                                                                                                                                                      & \begin{tabular}[c]{@{}c@{}}Need for effective information \\ \textbackslash{}sharing on threats\end{tabular}                                                             & \begin{tabular}[c]{@{}c@{}}Establishment of preventive actions \\ and countermeasures; \\ better detection and response\end{tabular}               & \begin{tabular}[c]{@{}c@{}}Malware Information Sharing \\ Platform (MISP)\end{tabular}                                                                \\ \hline
\cite{r13} & \begin{tabular}[c]{@{}c@{}}Paradigm shift in CTI exchange \\ with Ethereum \\ Blockchain Smart Contract \\ Marketplace\end{tabular}                                                               & Hesitation in voluntary CTI sharing                                                                                                                                      & \begin{tabular}[c]{@{}c@{}}Incentivized CTI sharing; \\ dynamic risk management \\ framework deployment\end{tabular}                               & \begin{tabular}[c]{@{}c@{}}Ethereum Blockchain, \\ W3C semantic web standards, \\ STIX\end{tabular}                                                   \\ \hline
\cite{r14} & \begin{tabular}[c]{@{}c@{}}Blockchain-based threat \\ intelligence sharing and \\ rating technology\end{tabular}                                                                                  & \begin{tabular}[c]{@{}c@{}}Tampering, lack of quality feedback, \\ and no incentive system for \\ providers\end{tabular}                                                 & \begin{tabular}[c]{@{}c@{}}Timely acquisition and analysis \\ of threat intelligence; \\ effective threat intelligence \\ ecosystem\end{tabular}   & \begin{tabular}[c]{@{}c@{}}Blockchain, \\ smart contracts\end{tabular}                                                                                \\ \hline
\cite{r15} & \begin{tabular}[c]{@{}c@{}}Taxonomy for classifying \\ threat-sharing technologies\end{tabular}                                                                                                   & Limitations of existing ontologies                                                                                                                                       & \begin{tabular}[c]{@{}c@{}}Identification of gaps and \\ differences in threat-sharing \\ technologies\end{tabular}                                & Classification of ontologies                                                                                                                          \\ \hline
\cite{r16} & \begin{tabular}[c]{@{}c@{}}BFLS: Blockchain and \\ Federated Learning for \\ CTI sharing\end{tabular}                                                                                             & \begin{tabular}[c]{@{}c@{}}Scalability, data privacy, and \\ security issues in CTI sharing\end{tabular}                                                                 & \begin{tabular}[c]{@{}c@{}}High accuracy threat detection; \\ secure CTI sharing\end{tabular}                                                      & \begin{tabular}[c]{@{}c@{}}Federated learning, \\ blockchain-based CTI \\ sharing platforms\end{tabular}                                              \\ \hline
\cite{r17} & \begin{tabular}[c]{@{}c@{}}ABC²: Awareness Architecture \\ Based on Blockchain CTI \\ Convergence\end{tabular}                                                                                    & Effective CTI sharing and evaluation                                                                                                                                     & \begin{tabular}[c]{@{}c@{}}Quality evaluation of CTI feeds; \\ trust-based reputation mechanism \\ for validators\end{tabular}                     & \begin{tabular}[c]{@{}c@{}}Proof-of-quality (PoQ) \\ consensus mechanism, \\ blockchain\end{tabular}                                                  \\ \hline
\cite{r18} & \begin{tabular}[c]{@{}c@{}}Collaborative CTI sharing \\ scheme using federated learning\end{tabular}                                                                                              & \begin{tabular}[c]{@{}c@{}}Privacy concerns and lack \\ of universal dataset formats\end{tabular}                                                                        & \begin{tabular}[c]{@{}c@{}}Robust ML-based network \\ intrusion detection system; \\ enhanced privacy and security\end{tabular}                    & Federated learning                                                                                                                                    \\ \hline
\cite{r19} & \begin{tabular}[c]{@{}c@{}}Cyber Threat Intelligence \\ Management Platform (CTIMP) \\ for industrial environments\end{tabular}                                                                   & \begin{tabular}[c]{@{}c@{}}Need for intelligent, \\ interoperable CTI sharing \\ technologies\end{tabular}                                                               & \begin{tabular}[c]{@{}c@{}}Advanced situational awareness; \\ cooperation, intelligent coping strategies, \\ and self-healing rules\end{tabular}   & CTIMP                                                                                                                                                 \\ \hline
\cite{r20} & \begin{tabular}[c]{@{}c@{}}Privacy-preserving architecture \\ for threat intelligence sharing \\ using Hyperledger Fabric and \\ MITRE ATT\&CK framework\end{tabular}                             & \begin{tabular}[c]{@{}c@{}}Inefficient, insecure, and \\ error-prone traditional threat \\ information sharing methods\end{tabular}                                      & \begin{tabular}[c]{@{}c@{}}Enhanced organizational security \\ and automation; \\ practical analysis of detected threats\end{tabular}              & \begin{tabular}[c]{@{}c@{}}Hyperledger Fabric, \\ MITRE ATT\&CK \\ framework\end{tabular}                                                             \\ \hline
\cite{r21} & \begin{tabular}[c]{@{}c@{}}Threat Intelligence Integrity \\ Audit (TIIA) scheme for IIoT\end{tabular}                                                                                             & \begin{tabular}[c]{@{}c@{}}Effective threat intelligence \\ sharing and information integrity \\ in IIoT\end{tabular}                                                    & \begin{tabular}[c]{@{}c@{}}Confidentiality, audit efficiency, \\ and reduced computational and \\ communication costs\end{tabular}                 & \begin{tabular}[c]{@{}c@{}}Lightweight Paillier \\ homomorphic encryption, \\ double chain structure\end{tabular}                                     \\ \hline
\cite{r22} & \begin{tabular}[c]{@{}c@{}}Blockchain-based framework \\ for differential CTI sharing\end{tabular}                                                                                                & \begin{tabular}[c]{@{}c@{}}Effective, flexible CTI sharing \\ with preserved information \\ integrity\end{tabular}                                                       & \begin{tabular}[c]{@{}c@{}}Trusted, verifiable, and \\ differential CTI sharing; \\ granularity and flexibility\end{tabular}                       & Ethereum private blockchain                                                                                                                           \\ \hline

\end{tabular}
\end{table*}

\begin{table*}[]
\setlength\tabcolsep{0pt}
\caption{Related work in Intelligence sharing using blockchain and DLT}
\begin{tabular}{|c|c|c|c|c|}
\hline
\textbf{Reference}          & \textbf{\begin{tabular}[c]{@{}c@{}}Proposed Work Methodology \\ with Algorithm Used\end{tabular}}                                                                                                 & \textbf{Problem Discussed}                                                                                                                                               & \textbf{Benefits and Outcomes}                                                                                                                     & \textbf{\begin{tabular}[c]{@{}c@{}}Application and Technology \\ Used\end{tabular}}                                                                   \\ \hline
\cite{r23} & \begin{tabular}[c]{@{}c@{}}DefenseChain with consortium \\ blockchain platform and \\ economic model\end{tabular}                                                                                 & \begin{tabular}[c]{@{}c@{}}Need for trustworthy, \\ cooperative defense mechanisms \\ against cyber threats in \\ \textbackslash{}cloud-hosted applications\end{tabular} & \begin{tabular}[c]{@{}c@{}}Effective collaboration in mitigating \\ cyber attacks, ensuring incentives \\ and trust\end{tabular}                   & \begin{tabular}[c]{@{}c@{}}Financial technology industry; \\ Hyperledger Composer\end{tabular}                                                        \\ \hline
\cite{r24} & \begin{tabular}[c]{@{}c@{}}Novel blockchain-based \\ architecture for CTI sharing\end{tabular}                                                                                                    & \begin{tabular}[c]{@{}c@{}}Challenges in CTI sharing: privacy, \\ trust, and accountability\end{tabular}                                                                 & \begin{tabular}[c]{@{}c@{}}Secure dissemination of CTI \\ data among organizations\end{tabular}                                                    & Blockchain technology                                                                                                                                 \\ \hline
\cite{r25} & \begin{tabular}[c]{@{}c@{}}Luunu platform with blockchain, \\ MISP, Model Cards, and \\ Federated Learning\end{tabular}                                                                           & \begin{tabular}[c]{@{}c@{}}Privacy, anonymity, and \\ security in CTI sharing\end{tabular}                                                                               & \begin{tabular}[c]{@{}c@{}}Enhanced transparency, \\ traceability, and \\ data provenance\end{tabular}                                             & \begin{tabular}[c]{@{}c@{}}Blockchain and federated \\ learning\end{tabular}                                                                          \\ \hline
\cite{r26} & \begin{tabular}[c]{@{}c@{}}Evaluation of blockchain \\ technology for CTI sharing in \\ IoT\end{tabular}                                                                                          & \begin{tabular}[c]{@{}c@{}}Centralized CTI sharing platform \\ limitations in IoT risk management\end{tabular}                                                           & \begin{tabular}[c]{@{}c@{}}Secure and efficient CTI \\ sharing\end{tabular}                                                                        & Blockchain technology                                                                                                                                 \\ \hline
\cite{r27} & \begin{tabular}[c]{@{}c@{}}Ethereum smart \\ contract-based CTI sharing\end{tabular}                                                                                                              & \begin{tabular}[c]{@{}c@{}}Confidential and anonymous CTI \\ sharing among financial institutions\end{tabular}                                                           & Secure information exchange                                                                                                                        & \begin{tabular}[c]{@{}c@{}}Ethereum blockchain \\ technology\end{tabular}                                                                             \\ \hline
\cite{r28} & \begin{tabular}[c]{@{}c@{}}Model with consortium blockchain \\ and distributed reputation \\ management systems; \\ Proof-of-Reputation (PoR) \\ consensus algorithm\end{tabular}                 & \begin{tabular}[c]{@{}c@{}}Byzantine attacks and unbalanced \\ performance in existing \\ blockchain-based models\end{tabular}                                           & \begin{tabular}[c]{@{}c@{}}Decentralized collaboration, \\ credible network environment\end{tabular}                                               & Consortium blockchain                                                                                                                                 \\ \hline
\cite{r29} & \begin{tabular}[c]{@{}c@{}}Blockchain-enabled framework \\ for CTI exchange in ICS\end{tabular}                                                                                                   & \begin{tabular}[c]{@{}c@{}}Privacy concerns and lack of \\ incentives in ICS security\end{tabular}                                                                       & \begin{tabular}[c]{@{}c@{}}Secure, private, and \\ incentivized CTI exchange\end{tabular}                                                          & Blockchain technology                                                                                                                                 \\ \hline
\cite{r30} & \begin{tabular}[c]{@{}c@{}}Incorporate Multi-Layer \\ Perceptron (MLP) within \\ blockchain nodes\end{tabular}                                                                                    & \begin{tabular}[c]{@{}c@{}}Challenges in fusing blockchain \\ technology and machine learning\end{tabular}                                                               & \begin{tabular}[c]{@{}c@{}}Intelligent, decentralized, and \\ tamper-proof network\end{tabular}                                                    & \begin{tabular}[c]{@{}c@{}}Blockchain and machine \\ learning\end{tabular}                                                                            \\ \hline
\cite{r31} & \begin{tabular}[c]{@{}c@{}}CITAShare, \\ consortium blockchain-based \\ threat intelligence sharing model\end{tabular}                                                                            & \begin{tabular}[c]{@{}c@{}}Privacy, trust, and sharing \\ mechanism issues in CTI sharing\end{tabular}                                                                   & \begin{tabular}[c]{@{}c@{}}Enhanced IT security through \\ efficient intelligence sharing\end{tabular}                                             & \begin{tabular}[c]{@{}c@{}}Consortium blockchain \\ technology\end{tabular}                                                                           \\ \hline
\cite{r32} & \begin{tabular}[c]{@{}c@{}}Decentralized platform using \\ EOS blockchain and IPFS \\ distributed hash table\end{tabular}                                                                         & \begin{tabular}[c]{@{}c@{}}Costs, risks, and legal reporting \\ requirements in CTI sharing\end{tabular}                                                                 & \begin{tabular}[c]{@{}c@{}}Incentivized information exchange, \\ support for legal \\ reporting requirements\end{tabular}                          & EOS blockchain and IPFS                                                                                                                               \\ \hline
\cite{r33} & \begin{tabular}[c]{@{}c@{}}Integration of distributed ledger \\ technologies (DLT) and a \\ generalized publish-subscribe \\ middleware with TAXII framework\end{tabular}                         & \begin{tabular}[c]{@{}c@{}}Secure and real-time exchange of \\ CTI data related to EPES \\ infrastructure security status\end{tabular}                                   & \begin{tabular}[c]{@{}c@{}}Data integrity, audit trail, and \\ near real-time CTI exchange\end{tabular}                                            & DLT and TAXII framework                                                                                                                               \\ \hline
\cite{r34} & \begin{tabular}[c]{@{}c@{}}Blockchain-based threat intelligence \\ sharing and rating technology\end{tabular}                                                                                     & \begin{tabular}[c]{@{}c@{}}Tampering, lack of feedback, and \\ absence of incentive systems in \\ CTI sharing\end{tabular}                                               & \begin{tabular}[c]{@{}c@{}}Efficient protection and \\ emergency response\end{tabular}                                                             & \begin{tabular}[c]{@{}c@{}}Blockchain technology, \\ smart contracts\end{tabular}                                                                     \\ \hline
\cite{r35} & \begin{tabular}[c]{@{}c@{}}DefenseChain with permissioned \\ blockchain architecture, \\ reputation system, and \\ economic model\end{tabular}                                                    & \begin{tabular}[c]{@{}c@{}}Challenges in threat intelligence \\ sharing\end{tabular}                                                                                     & \begin{tabular}[c]{@{}c@{}}Improved performance, \\ benefits of cooperative \\ real-time threat intelligence \\ sharing\end{tabular}               & \begin{tabular}[c]{@{}c@{}}Permissioned blockchain, \\ Hyperledger Composer\end{tabular}                                                              \\ \hline
\cite{r36} & \begin{tabular}[c]{@{}c@{}}Decentralized infrastructure for CTI \\ sharing with controlled access, \\ authentication, and \\ SDN control plane\end{tabular}                                       & \begin{tabular}[c]{@{}c@{}}Common issues in threat \\ intelligence sharing\end{tabular}                                                                                  & \begin{tabular}[c]{@{}c@{}}Secure, scalable, \\ cost-effective platform; \\ fast security policy enforcement\end{tabular}                          & \begin{tabular}[c]{@{}c@{}}Smart contracts, \\ Software-Defined Networking \\ (SDN)\end{tabular}                                                      \\ \hline
\cite{r37} & \begin{tabular}[c]{@{}c@{}}Blockchain-based network \\ threat intelligence sharing platform\end{tabular}                                                                                          & \begin{tabular}[c]{@{}c@{}}Isolated information silos \\ limiting data sharing\end{tabular}                                                                              & \begin{tabular}[c]{@{}c@{}}Secure and private collection \\ of diverse, \\ large-scale network data\end{tabular}                                   & Blockchain technology                                                                                                                                 \\ \hline
\cite{r38} & \begin{tabular}[c]{@{}c@{}}Blockchain-based open CTI\\  framework with traceability, \\ integrity, and Sybil-resistance\end{tabular}                                                              & \begin{tabular}[c]{@{}c@{}}Collecting large amounts of accurate\\ , non-malicious data for \\ analysis and sharing\end{tabular}                                          & \begin{tabular}[c]{@{}c@{}}Prevention of Sybil attacks \\ and blocking of \\ malicious data injection\end{tabular}                                 & Blockchain technology                                                                                                                                 \\ \hline
\cite{r39} & \begin{tabular}[c]{@{}c@{}}Blockchain-based architecture \\ with reputation levels and \\ topic-based independent ledgers\end{tabular}                                                            & \begin{tabular}[c]{@{}c@{}}Trusted environment for sharing \\ cyber-intelligence information\end{tabular}                                                                & \begin{tabular}[c]{@{}c@{}}Integrity, privacy, \\ confidentiality, and \\ truthfulness of shared information\end{tabular}                          & Blockchain technology                                                                                                                                 \\ \hline
\cite{r40} & \begin{tabular}[c]{@{}c@{}}TITAN, a trust enhancement framework \\ using P2P reputation systems, \\ \textbackslash{}blockchain, and \\ \textbackslash{}Trusted Execution Environment\end{tabular} & Trust issues in decentralized sharing                                                                                                                                    & \begin{tabular}[c]{@{}c@{}}Security, integrity, and \\ privacy enhancement\end{tabular}                                                            & \begin{tabular}[c]{@{}c@{}}Blockchain and \\ Trusted Execution Environment \\ technologies\end{tabular}                                               \\ \hline
\cite{r41} & \begin{tabular}[c]{@{}c@{}}Healthcare Data Gateway (HGD) \\ app using blockchain technology\end{tabular}                                                                                          & \begin{tabular}[c]{@{}c@{}}Privacy risks and data sharing \\ challenges in healthcare\end{tabular}                                                                       & \begin{tabular}[c]{@{}c@{}}Secure and private patient data \\ sharing, improved intelligence \\ of healthcare systems\end{tabular}                 & Blockchain technology                                                                                                                                 \\ \hline
\end{tabular}
\end{table*}

The rapid development of computer and network technology has led to frequent cyber security incidents and numerous new vulnerabilities, highlighting the importance of threat intelligence. However, existing sharing mechanisms are susceptible to tampering, lack quality feedback, and have no incentive system for providers. The paper \cite{r14} proposes a blockchain-based threat intelligence sharing and rating technology to address these issues. By leveraging the properties of blockchain, such as openness, consensus, autonomy, decentralization, trustlessness, non-tampering, and traceability, the authors construct blocks containing various threat intelligence information. They design a threat intelligence sharing and rating system based on blockchain, introducing corresponding sharing methods, rating methods, and smart contracts. This approach enables timely acquisition and analysis of valuable threat intelligence information, fostering a continuously effective threat intelligence ecosystem. The paper also provides an experimental environment and smart contract design to demonstrate the effectiveness of the proposed solution.

The current cyber threat intelligence information exchange ecosystem relies on automation to effectively share threat intelligence. However, existing ontologies, such as OpenIOC, STIX, and IODEF, are often based on use cases, which may not always be relevant for future threats. This episodic approach can lead to the exclusion of valuable information \cite{r15}. To address this limitation, this paper proposes a taxonomy for classifying threat-sharing technologies. This agnostic framework aims to classify existing technologies, identify gaps, and elucidate their differences from a scientific perspective. The authors are also working on developing a thesaurus to describe, compare, and classify detailed cyber security terms, focusing on the classification of the ontologies themselves \cite{t1}.

The increasing number of cyber attacks necessitates an effective approach to cyber threat intelligence (CTI) sharing, while maintaining data privacy and security. This study introduces a novel method, BFLS (Blockchain and Federated Learning for sharing threat detection models as Cyber Threat Intelligence), which leverages federated learning for scalable machine learning applications and blockchain-based CTI sharing platforms for enhanced security and privacy \cite{r16}, \cite{t3}. The consensus protocol within the blockchain is refined to select high-quality CTIs for federated learning, with models automatically aggregated and updated via smart contracts. Experimental results on ISCX-IDS-2012 and CIC-DDoS-2019 datasets demonstrate BFLS's high accuracy in threat detection and its ability to securely share CTI.

The problem of effectively sharing and evaluating cyber threat intelligence (CTI) while maintaining data security is of great importance. The research work in \cite{r17} proposes a blockchain-based system called ABC² (Awareness Architecture Based on Blockchain CTI Convergence) that focuses on CTI sharing using blockchain technology and a novel consensus mechanism called proof-of-quality (PoQ). ABC² aims to evaluate the quality of CTI feeds and contextualize the reputation of CTI sources based on quality parameters, utilizing a trust-based reputation mechanism for selecting validators. The PoQ mechanism ensures a transparent and secure evaluation process, creating a reliable and distributed repository of CTI feeds.

Designing an ML-based network intrusion detection system using heterogeneous data from different sources and organizations poses challenges due to privacy concerns and the lack of universal dataset formats. Authors proposes a collaborative cyber threat intelligence sharing scheme using federated learning, which allows multiple organizations to jointly develop a robust ML-based network intrusion detection system without sharing sensitive user data \cite{r18}. The proposed framework demonstrates its effectiveness by classifying various traffic types originating from multiple organizations without inter-organizational data exchange, thus enhancing the security and privacy of threat intelligence sharing \cite{t2}.

Addressing modern cyber threats requires intelligent and interoperable Cyber Threat Information (CTI) sharing technologies that ensure high-quality, organized, and comprehensible data. Authors in \cite{r19} introduces an innovative Cyber Threat Intelligence Management Platform (CTIMP) for industrial environments, which combines trusted public source information with relevant internal organizational data. The platform's advanced visualization mechanism and user interface enhance situational awareness and enable extended cooperation, intelligent coping strategy selection, and automated self-healing rules for addressing cyber threats effectively.

Traditional threat information sharing methods can be inefficient, insecure, and error-prone. This paper proposes a privacy-preserving, trustworthy architecture for threat intelligence sharing based on permissioned blockchain technology, specifically Hyperledger Fabric, and the MITRE ATT\&CK framework. The contributions include the development of a new framework, practical analysis of detected threats for generating meaningful reports, a proof-of-concept implementation with security analysis, and performance measurements to validate the feasibility and effectiveness of the proposed solution in enhancing organizational security and automation \cite{r20}.

The growing cyber security threats in Industrial Internet of Things (IIoT) systems necessitate effective threat intelligence sharing while maintaining information integrity and building a complete attack chain. A blockchain-enabled Threat Intelligence Integrity Audit (TIIA) scheme for IIoT, featuring a double chain structure has been proposed in \cite{r21}. The TIIA scheme employs lightweight Paillier homomorphic encryption to ensure confidentiality during the sharing process and includes an audit scheme based on lightweight technology. Additionally, a fast deletion algorithm of redundant blocks is designed to improve audit efficiency and reduce computational and communication costs \cite{t4}.

The increasing cyber security threats in IIoT systems call for more effective and flexible threat intelligence sharing while preserving information integrity. A blockchain-based framework has been proposed in \cite{r22} that enables differential sharing of Cyber Threat Intelligence (CTI) using policies/metrics defined by CTI producers, without compromising verifiability for CTI consumers. Key contributions include the concept of differential sharing in the CTI context, a detailed design of the proposed framework, and a proof-of-concept implementation using Ethereum private blockchain. This approach offers more granularity and flexibility compared to existing solutions, providing trusted, verifiable, and differential CTI sharing.

The need for trustworthy, cooperative defense mechanisms against cyber threats in cloud-hosted applications has become essential. The authors in \cite{r23} presents "DefenseChain," a novel threat intelligence sharing and defense system using a consortium blockchain platform and an economic model. DefenseChain enables organizations to collaborate in mitigating the impact of cyber attacks while ensuring incentives and trust. Applied in the financial technology industry, DefenseChain demonstrates its effectiveness in a real-world insurance claim processing use case. Experimental results on an Open Cloud testbed using Hyperledger Composer show that DefenseChain outperforms existing solutions in selecting appropriate detector and mitigator peers, ultimately mitigating threat risk.

The growing number of cyberattacks has highlighted the importance of Cyber Threat Intelligence (CTI) sharing, which faces challenges like privacy, trust, and accountability. Authors in has been proposed in \cite{r24} introduces a novel blockchain-based architecture designed to securely disseminate CTI data among organizations. By leveraging key blockchain features such as decentralization, cryptographic keys, and immutability, the proposed architecture addresses the issues of trust, privacy, and accountability. The study provides a comprehensive analysis of existing blockchain-based CTI sharing proposals and presents a detailed architectural design for a democratically anonymous and trusted CTI sharing system, demonstrating its suitability for practical, real-world environments.

The sensitive nature of Cyber Threat Intelligence (CTI) sharing demands a system that ensures privacy, anonymity, and security for participating organizations. Authors in has been proposed in \cite{r25} introduces "Luunu," a blockchain, MISP, Model Cards, and Federated Learning-enabled CTI sharing platform that addresses these challenges while providing enhanced transparency, traceability, and data provenance. Luunu incorporates self-sovereign identity to maintain participant anonymity and employs a blockchain-based federated learning system to analyze collected CTI data. The platform's main contributions include a blockchain-based CTI sharing platform, enhanced transparency and provenance with MISP Model Card objects, a coordinator-less federated machine learning approach for CTI analysis, and self-sovereign identity-enabled mobile wallets for anonymous reporting \cite{t5}.

The growing IoT landscape necessitates improved risk management for organizational infrastructure, with existing centralized CTI sharing platforms falling short. The research work in has been proposed in \cite{r26} evaluates blockchain technology's potential to overcome these limitations by addressing CTI sharing challenges securely and efficiently. We explore blockchain's opportunities, discuss relevant research, and highlight unique future research questions in the context of distributed intelligence sharing.

Current informal CTI sharing among organizations is highly subjective and dependent on individual social networks. To confidentially and anonymously share valuable intelligence among financial institutions, this research proposes a new method using Ethereum smart contract blockchain technology. By hashing device identity and replacing it with an on-chain verifiable random function, the identity of participating nodes is protected, ensuring secure information exchange \cite{r27}.

CTI sharing enhances cybersecurity responsiveness, but existing blockchain-based models face challenges like byzantine attacks and unbalanced performance. Authors introduces a new model combining consortium blockchain and distributed reputation management systems to automate tactical threat intelligence sharing has been proposed in \cite{r28}. The proposed "Proof-of-Reputation" (PoR) consensus algorithm meets transaction rate requirements while maintaining a credible network environment. Key contributions include: (1) a decentralized collaboration consortium automating CTI sharing while addressing security concerns, and (2) the PoR consensus algorithm and reputation model, reducing the impact of byzantine behaviors in the CTI sharing collaboration consortium.

ICS security is crucial, but organizations are hesitant to share CTI due to privacy concerns and lack of incentives. This paper presents a novel blockchain-enabled framework to facilitate secure, private, and incentivized CTI exchange related to ICS. Key contributions include: A comprehensive review of existing CTI-sharing solutions, highlighting critical issues, a blockchain-enabled CTI sharing framework for ICS, offering incentives to encourage information exchange and a complete system design, with use-case scenarios demonstrating the framework's suitability in real-world applications, addressing privacy, trust, and security concerns \cite{r29}.

The fusion of blockchain technology and machine learning presents notable challenges, yet it promises substantial advantages, including the establishment of an intelligent, decentralized, and tamper-proof network. Authors in \cite{r30} makes several contributions to this field by proposing an efficient method to incorporate a Multi-Layer Perceptron (MLP) model within each blockchain network node, minimizing the processing power and time needed to develop an intelligent blockchain network. Additionally, the paper ensures that every node possesses knowledge of the model architecture during the network formation process. It achieves this by training a randomly selected node's model and subsequently replicating the intelligence across the entire network. This work demonstrates the promising potential of merging blockchain and machine learning technologies to create a novel paradigm.

The increasing complexity of cyber threats, such as Advanced Persistent Threat (APT) attacks, demands more efficient intelligence sharing among organizations. However, privacy, trust, and sharing mechanism issues often hinder the process. In response to these challenges, \cite{r31} introduces CITAShare, a new threat intelligence sharing model based on the consortium blockchain technology. CITAShare includes a distributed architecture database and relies on consensus algorithms for data updates. The model utilizes smart contracts to facilitate the sharing of threat intelligence, addressing privacy concerns in the process. Additionally, to encourage participation in intelligence sharing, an incentive mechanism based on an improved Shapley value is proposed for profit distribution. This approach ensures operational rationality by employing smart contracts in the specific distribution process.

The exchange of threat intelligence is crucial for enhancing IT security but often faces challenges due to costs, risks, and legal reporting requirements. Existing platforms lack incentives and fail to address these reporting obligations. Authors presents a decentralized threat intelligence sharing platform that supports legal reporting requirements while offering incentives for information exchange \cite{r32}. The platform, implemented using the EOS blockchain and IPFS distributed hash table, ensures availability, integrity, and non-repudiation through its distributed ledger technology (DLT). Furthermore, the platform utilizes blockchain tokens to assign real value to threat intelligence, providing decentralized incentives. Our prototype and cost measurements demonstrate the feasibility and cost-efficiency of the proposed concept.

Authors in \cite{r33} addresses the challenges of secure and real-time exchange of cyber threat intelligence (CTI) data related to the security status of EPES infrastructures by enhancing the TAXII framework. We propose a novel approach that integrates distributed ledger technologies (DLT) and a generalized publish-subscribe middleware to ensure data integrity, audit trail, and near real-time CTI exchange. The combination of TAXII framework and DLTs provides a secure, tamper-proof, and highly scalable solution for information sharing. The applicability of our proposed solution is verified through a series of experiments conducted on the target prototype.

The current threat intelligence sharing mechanisms face issues with tampering, lack of feedback, and absence of incentive systems for providers. There is an urgent need to address these problems while also evaluating the quality, credibility, and contribution rates of threat intelligence sources for efficient protection and emergency response. The research work in \cite{r34} proposes a blockchain-based threat intelligence sharing and rating technology to tackle these challenges. Leveraging the unique properties of blockchain, such as decentralization, trustlessness, and traceability, the system designs a threat intelligence sharing and rating process using smart contracts. This approach enables timely and effective acquisition and analysis of valuable threat intelligence information, promoting the continuous development of the threat intelligence ecosystem. The experimental setup and smart contract design demonstrate the effectiveness of this technology, which has broad market applications across various industries, including cybersecurity, finance, government, industrial internet, and 5G communication operators.

Addressing the challenges of threat intelligence sharing, this paper introduces a novel "DefenseChain" platform for two-stage cyber defense using a permissioned blockchain architecture. The approach \cite{r35} offers shorter deployment times and reduced resource-intensive properties, making it suitable for secure data sharing among a federation of organizations. The platform also includes a reputation system and protocols that objectively rate peers based on 'Quality of Detection' and 'Quality of Mitigation' metrics. An economic model is proposed to ensure consortium sustainability and discourage false reporting or free-riding. Implemented using Hyperledger Composer in an NSF Cloud testbed, the DefenseChain platform demonstrates improved performance compared to existing solutions, showcasing the benefits of cooperative real-time threat intelligence sharing.

The work in \cite{r36} aims to address common issues in threat intelligence sharing by proposing a secure, scalable, and cost-effective decentralized infrastructure for various parties to share cyber threat intelligence. The platform offers high security, enabling members to share sensitive information through controlled access and authentication. It also ensures trustworthiness, as participants benefit from a reliable business model. This is a necessary prerequisite for a successful security collaboration based on smart contracts for providing reliable SLAs. Lastly, the platform incorporates a Software-Defined Networking (SDN) control plane that enables fast security policy enforcement, ultimately reducing cyber attack mitigation time.

The growing severity of cyber threats and the variety of attack methods necessitate the development of an efficient cyber threat intelligence ecosystem. Enhancing collaboration and interconnectivity among information systems is crucial for maximizing threat intelligence value and improving threat detection and emergency response capabilities. However, existing approaches often result in isolated information silos, limiting effective data sharing. The work done in \cite{r37} proposes a blockchain-based network threat intelligence sharing platform that addresses these challenges. Experimental results demonstrate that the platform can securely and privately collect diverse, large-scale network data, significantly improving the efficiency of sharing network threat intelligence across organizations.

The challenge of collecting large amounts of accurate and non-malicious data to analyze and share is faced by CTI systems. To address this, a blockchain-based open CTI framework is proposed that provides traceability, integrity, and Sybil-resistance as shown in Figure. The framework in \cite{r38}consists of contributors who collect and share threat-related data, consumers who consume such data, and feeds that provide CTI data sharing services. It allows data collection through contributors to maximize the ability to collect threat-related data while preventing Sybil attacks from malicious contributors. The data verification performed by the CTI feed also degrades the data dissemination capability of malicious contributors, allowing the CTI system to block malicious data injection automatically.

\begin{figure*}[!ht]
\centering
\includegraphics[width=15cm,height=13cm]{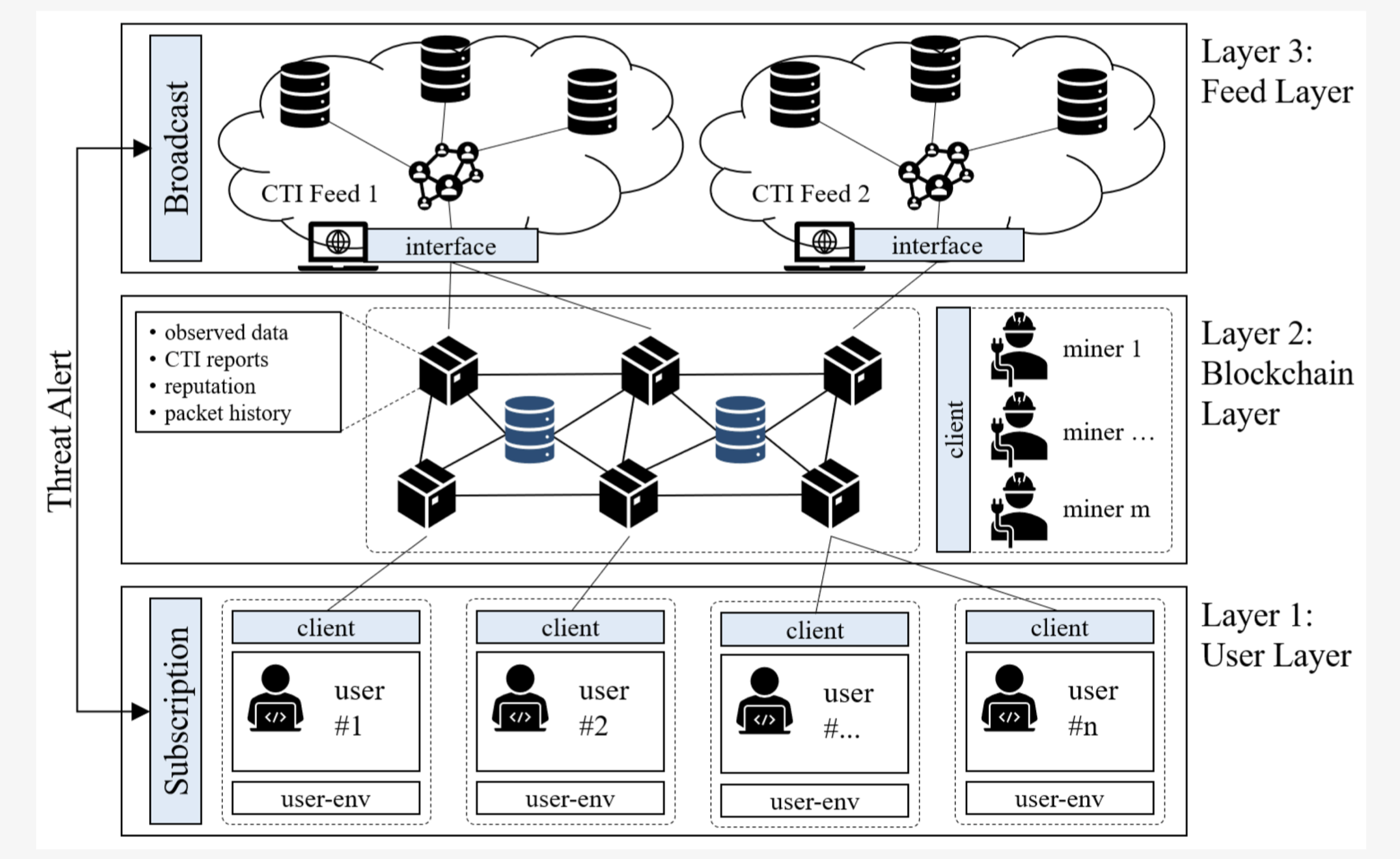}
\caption{Illustration intelligence sharing framework discussed in \cite{r38}}
\label{fog}
\end{figure*}

Sharing cyber-intelligence information is crucial for organizations to enhance their security plans and teams, making them more resilient against cyberattacks. However, information sharing requires a trusted environment that guarantees the integrity, privacy, confidentiality, and truthfulness of the information shared. Authors in \cite{r39} proposes a blockchain-based architecture that assigns reputation levels to each participant and credits them based on the accuracy of the validation they provide. The architecture also organizes information into topics and instantiates them in independent ledgers to ensure their security. The proposed architecture was validated in a proof-of-concept scenario involving three organizations.

The challenges associated with cyber threat intelligence sharing include privacy concerns, policy/legal issues, negative publicity, and the high cost of sharing \cite{t7}. While decentralized blockchain-based sharing architectures have been developed to address these challenges, issues related to trust remain unsolved. To address this issue, this paper proposes a trust enhancement framework, called TITAN, that uses P2P reputation systems to enhance trust in decentralized sharing \cite{r40}. The framework uses blockchain and Trusted Execution Environment technologies to ensure security, integrity, and privacy. The paper discusses the design and progress of the framework and identifies the remaining challenges to be addressed.

The sharing of healthcare data is crucial for improving the quality of healthcare services, but currently, patient data is scattered across various healthcare systems, posing privacy risks and hindering data sharing. The paper \cite{r41} proposes a solution to this problem by introducing an App, the Healthcare Data Gateway (HGD), that uses blockchain technology to allow patients to own, control and share their data securely without compromising their privacy. This architecture provides a new way to improve the intelligence of healthcare systems while maintaining the confidentiality of patient data, similar to how blockchain has been implemented in the financial sector for auditable computing using a decentralized network of peers and a public ledger \cite{t6}.

\section{National Cybersecurity Strategy and Its Implications}
\subsection{Existence of a National Cybersecurity Strategy}

A National Cybersecurity Strategy (NCS) plays a crucial role in addressing cybersecurity risks and outlining solutions to tackle these challenges. The development of an NCS demonstrates a government's commitment to protect its digital infrastructure, the privacy of its citizens, and maintain national security \cite{61}. NCSs vary among countries, reflecting their unique contexts and priorities. However, they share common objectives, such as securing critical infrastructure, promoting cybersecurity awareness, fostering public-private partnerships, and establishing legal frameworks for tackling cybercrimes \cite{62}.

The United States' NCS, for example, highlights the importance of information sharing between the public and private sectors \cite{63}. The strategy emphasizes the need to improve the detection and prevention of cyber threats through effective intelligence sharing. Similarly, the United Kingdom's NCS underscores the significance of collaboration, innovation, and strong governance in enhancing the country's cybersecurity posture \cite{64}. The strategies from both countries stress the importance of establishing mechanisms that facilitate intelligence sharing while also addressing privacy concerns. An NCS also serves as a vital tool for fostering cooperation among different stakeholders, including government agencies, private organizations, and international partners. This cooperation enables countries to leverage collective resources and expertise in addressing cyber threats and ensuring a more resilient digital ecosystem \cite{65}. As the cyber landscape evolves, national strategies must adapt and respond to emerging trends and challenges. Regular reviews and revisions of NCSs are crucial to ensure that they remain relevant and effective in addressing current and future cyber threats \cite{66}.

One notable example of a country adapting its NCS is Estonia, which has become a global leader in cybersecurity following a series of cyberattacks in 2007. Estonia's NCS emphasizes the importance of a proactive approach to cybersecurity, focusing on the development of a strong cyber defense and building a resilient digital infrastructure. The Estonian NCS also prioritizes public awareness and education, recognizing the need to foster a cybersecurity-conscious culture within the country \cite{67}. Another example is Singapore, which has developed a holistic NCS that integrates multiple domains, such as critical information infrastructure protection, research and development, and international engagement. Singapore's approach highlights the importance of public-private partnerships in driving innovation and fostering a secure and trusted digital environment \cite{68}.

\subsection{Addressing Cybersecurity Risks and Solutions}

Intelligence sharing has emerged as an essential component of national cybersecurity strategies, as it helps nations better understand and respond to cyber threats \cite{69}. The implementation of Blockchain and Distributed Ledger Technology (DLT) in this context can enhance the effectiveness of intelligence sharing while mitigating potential risks. These technologies can offer various benefits, such as improved data confidentiality, integrity, and availability \cite{611}. 

Several scientific papers have explored the use of Blockchain and DLT to address cybersecurity risks. For instance, \cite{612} propose the use of Blockchain to enable secure and efficient information sharing among organizations, while \cite{613} suggest that Blockchain can be leveraged to improve the security and privacy of the Internet of Things. Furthermore, \cite{614} assert that DLT can be employed to strengthen the trustworthiness of digital identities, which is critical for secure intelligence sharing.

However, implementing Blockchain and DLT for intelligence sharing is not without challenges. Privacy concerns arise due to the transparent and immutable nature of these technologies \cite{615}. Moreover, the relatively low transaction throughput of some Blockchain platforms may hinder their ability to support real-time intelligence sharing \cite{616}. To address these challenges, researchers have proposed various solutions, such as the integration of privacy-preserving mechanisms like zero-knowledge proofs \cite{617} and the development of scalable Blockchain architectures \cite{618}.

A study in \cite{619} investigates the potential of Blockchain for securing electronic health records (EHRs) by addressing data integrity and privacy issues. The authors propose a model that utilizes Blockchain and access control mechanisms to ensure the confidentiality and security of EHRs. This study highlights the potential of Blockchain technology to improve security in highly sensitive industries.

Another important research paper \cite{620} explores the challenges and potential of utilizing Blockchain technology to secure critical infrastructure. The authors examine the requirements for a secure and resilient critical infrastructure, and how Blockchain technology can address these needs. The paper discusses the potential of Blockchain to provide a robust, distributed, and tamper-proof platform for managing the security of critical infrastructure.

Governments should also consider the international dimensions of cybersecurity and the role of cross-border cooperation in addressing global cyber threats. Collaboration between countries can be facilitated by sharing best practices, establishing common cybersecurity standards, and launching joint initiatives. The role of regional and international organizations, such as the European Union, NATO, and the United Nations, can be instrumental in facilitating this cooperation and ensuring a harmonized approach to cybersecurity \cite{621}.

Moreover, building a strong cybersecurity workforce is a vital component of any NCS. Governments must invest in education and training programs that equip individuals with the necessary skills to address the cybersecurity challenges of the digital age. This may include establishing cybersecurity-focused academic programs, encouraging professional certifications, and offering incentives for skilled professionals to enter the cybersecurity field \cite{622}).

In addition to workforce development, national cybersecurity strategies should also promote research and development in cybersecurity technologies. Governments can support such efforts by providing funding, establishing research centers, and collaborating with private sector organizations and academic institutions. Investments in research and development can lead to the discovery of innovative solutions and contribute to the overall resilience of a nation's digital infrastructure \cite{623}.

In conclusion, a comprehensive understanding of existing NCSs is essential for organizations seeking to address cybersecurity problems. These strategies can provide valuable insights into the potential role of Blockchain and DLT in enhancing intelligence sharing, as well as the challenges and implications associated with their implementation. By analyzing NCSs, organizations can determine whether the proposed solutions align with their specific needs and objectives.

Furthermore, this analysis may identify any gaps in the strategy that need to be addressed to effectively mitigate the risks associated with the problem. Thus, a comprehensive understanding of the National Cybersecurity Strategy is essential for any organization seeking to address the cybersecurity problem. By addressing the aforementioned question, organizations can build a comprehensive understanding of existing cybersecurity strategies and their potential impact on solving the problem.

\section{implementation of Blockchain and Distributed Ledger Technology (DLT) for intelligent sharing}
Blockchain and Distributed Ledger Technology (DLT) are increasingly being adopted for various applications due to their inherent security, transparency, and decentralization. One such application is intelligent sharing, which involves the secure exchange of information between various entities in a network. In this article, we discuss the experimental setup, available datasets, and suitable metrics for implementing and evaluating Blockchain and DLT solutions for intelligent sharing.

\subsection{Experimental Setup} 
Experimental Setup for Implementing Blockchain and DLT for Intelligence Sharing
To implement Blockchain and DLT for intelligence sharing, researchers must create a suitable experimental setup that addresses the following aspects:

\subsubsection{Network Architecture}

Researchers should establish a network architecture that allows for the secure exchange of information between various nodes in the system. This includes the deployment of smart contracts to automate transactions and enforce data sharing policies, as well as the use of consensus algorithms such as Proof of Work, Proof of Stake, or other Byzantine Fault Tolerant mechanisms.

\subsubsection{Security and Privacy}

Ensuring security and privacy in the proposed system is crucial, particularly for sensitive data sharing applications. Researchers should implement advanced cryptographic techniques such as zero-knowledge proofs, homomorphic encryption, or secure multi-party computation to protect data privacy while maintaining data integrity.

\subsubsection{Scalability and Performance}

The experimental setup should be capable of handling a large volume of transactions and maintaining high throughput for effective data sharing (Wang et al., 2021). Researchers must also consider the trade-offs between security, performance, and decentralization while designing their systems.

\subsection{Datasets for Intelligent Sharing}

\begin{table*}[]
\setlength\tabcolsep{0pt}
\caption{Dataset used for BC and DLT implementation for intelligence sharing}
\begin{tabular}{|c|c|c|c|c|}
\hline
\textbf{Data Set Name}                                                        & \textbf{Benefit of Use}                                                                                                   & \textbf{Type of Data}                                                                                     & \textbf{Data Access}                                                                                    & \textbf{Application Domain}                                                                  \\ \hline
\begin{tabular}[c]{@{}c@{}}The Honeynet Project's \\ Shared Data\end{tabular} & \begin{tabular}[c]{@{}c@{}}Provides data from honeypots for \\ cybersecurity research and development\end{tabular}        & \begin{tabular}[c]{@{}c@{}}Honeypot data, malware samples, \\ network traffic, attack logs\end{tabular}   & Publicly available                                                                                      & \begin{tabular}[c]{@{}c@{}}Cybersecurity, \\ network security\end{tabular}                   \\ \hline
IMPACT Datasets                                                               & \begin{tabular}[c]{@{}c@{}}Offers a wide range of \\ cybersecurity research data\end{tabular}                             & \begin{tabular}[c]{@{}c@{}}Network traffic, malware, \\ intrusion detection, vulnerabilities\end{tabular} & \begin{tabular}[c]{@{}c@{}}Publicly available, \\ Registration required\end{tabular}                    & \begin{tabular}[c]{@{}c@{}}Cybersecurity, network security, \\ malware analysis\end{tabular} \\ \hline
\begin{tabular}[c]{@{}c@{}}VERIS Community \\ Database\end{tabular}           & \begin{tabular}[c]{@{}c@{}}Enables access to a repository of \\ cybersecurity incidents\end{tabular}                      & Cybersecurity incident reports                                                                            & Publicly available                                                                                      & Cybersecurity, incident response                                                             \\ \hline
MISP                                                                          & \begin{tabular}[c]{@{}c@{}}Facilitates sharing, storing, and \\ correlating IOCs and threat intelligence\end{tabular}     & \begin{tabular}[c]{@{}c@{}}Indicators of compromise (IOCs), \\ threat intelligence\end{tabular}           & \begin{tabular}[c]{@{}c@{}}Publicly available, \\ Registration required\end{tabular}                    & \begin{tabular}[c]{@{}c@{}}Cyber threat intelligence, \\ malware analysis\end{tabular}       \\ \hline
CVE Database                                                                  & \begin{tabular}[c]{@{}c@{}}Provides standardized information \\ about security vulnerabilities and exposures\end{tabular} & \begin{tabular}[c]{@{}c@{}}Software vulnerabilities \\ and exposures\end{tabular}                         & Publicly available                                                                                      & Cybersecurity, software security                                                             \\ \hline
CTIIC Datasets                                                                & \begin{tabular}[c]{@{}c@{}}Offers a range of cyber threat intelligence \\ datasets from the U.S. government\end{tabular}  & \begin{tabular}[c]{@{}c@{}}Cyber threat intelligence, \\ national security incidents\end{tabular}         & \begin{tabular}[c]{@{}c@{}}Publicly available, some datasets \\ may have restricted access\end{tabular} & \begin{tabular}[c]{@{}c@{}}Cyber threat intelligence, \\ national security\end{tabular}      \\ \hline
\end{tabular}
\end{table*}
To evaluate the effectiveness of Blockchain and DLT for intelligent sharing, researchers can use various available datasets, depending on the specific application area. Some of the widely used datasets include:

Datasets such as the Credit Card Fraud Detection dataset from the Machine Learning Repository can be used to study Blockchain and DLT solutions for secure financial transactions. This dataset contains 284,807 transactions, with 492 fraudulent transactions. The MIMIC-III dataset, a comprehensive and publicly available dataset containing electronic health records from over 40,000 patients, can be used to evaluate the effectiveness of Blockchain and DLT solutions for secure healthcare data sharing. Datasets such as the IoT Network Intrusion Dataset from the Canadian Institute for Cybersecurity can be employed to study the effectiveness of Blockchain and DLT in securing IoT networks (Alrawais et al., 2017). This dataset contains approximately 757,000 records of network traffic in IoT devices.

Datasets play a crucial role in the development and evaluation of intelligent sharing solutions, including those focused on cyber threat intelligence sharing. These datasets enable researchers to analyze the effectiveness of Blockchain and DLT solutions in various application domains, such as cybersecurity. Here are some relevant datasets for intelligent sharing and cyber threat intelligence sharing:

The Honeynet Project is a non-profit \cite{36}, global organization that focuses on the research and development of cybersecurity tools and techniques. They provide various datasets collected from honeypots and other security-related systems, which can be used to study the effectiveness of Blockchain and DLT solutions for cyber threat intelligence sharing.

The Information Marketplace for Policy and Analysis of Cyber-risk \& Trust (IMPACT) Datasets \cite{37} is a collaborative effort between the U.S. Department of Homeland Security and the National Science Foundation to facilitate the access and sharing of cybersecurity research data. They provide a wide range of datasets related to cybersecurity, including network traffic, malware, and intrusion detection, which can be utilized for studying the implementation of Blockchain and DLT for secure cyber threat intelligence sharing.

The VERIS Community Database (Vocabulary for Event Recording and Incident Sharing) Community Database is an open-source repository of cybersecurity incidents, contributed by organizations worldwide \cite{38}. This dataset can be used to study the effectiveness of Blockchain and DLT solutions in securely sharing cyber threat intelligence information and incident reports.

The Malware Information Sharing Platform (MISP) \cite{39} is an open-source platform for sharing, storing, and correlating indicators of compromise (IOCs) and threat intelligence. It provides a rich dataset of threat intelligence data that can be used to study the efficacy of Blockchain and DLT solutions in secure cyber threat intelligence sharing.

The Common Vulnerabilities and Exposures (CVE) database \cite{40} is a widely-used, publicly available repository of standardized information about security vulnerabilities and exposures. This dataset can be employed to evaluate the effectiveness of Blockchain and DLT solutions in securely sharing information about software vulnerabilities and other cyber threats.

The Cyber Threat Intelligence Integration Center (CTIIC) Datasets \cite{41}, part of the U.S. government's intelligence community, provides a range of datasets related to cyber threat intelligence. Researchers can use these datasets to study the implementation and effectiveness of Blockchain and DLT solutions for secure cyber threat intelligence sharing.

These datasets serve as valuable resources for researchers working on intelligent sharing and cyber threat intelligence sharing applications using Blockchain and DLT. By leveraging these datasets, researchers can gain insights into the potential benefits and limitations of using Blockchain and DLT technologies for secure, efficient, and scalable cyber threat intelligence sharing.

\subsection{Metrics for Evaluating Blockchain and DLT Solutions for Intelligent Sharing}
Several metrics can be used to measure the accuracy of Blockchain and DLT solutions for intelligent sharing. These include:
\subsubsection{False Positive, True Positive, True Negative, and False Negative Rates}

These metrics measure the performance of a classification algorithm in correctly identifying and classifying data points. They are crucial for evaluating the effectiveness of Blockchain and DLT solutions in identifying and handling secure transactions or data sharing events.

\subsubsection{F1 Score and F2 Score}

The F1 and F2 scores are harmonic means of precision and recall, with the F2 score assigning more weight to recall. These metrics can be used to evaluate the trade-offs between false positives and false negatives in the proposed Blockchain and DLT solutions.

\subsubsection{Area Under the Curve (AUC) and Receiver Operating Characteristic (ROC) Curve}

The AUC and ROC curve can be used to measure the overall performance of a classification algorithm across different classification thresholds. A higher AUC value indicates better classification performance, while the ROC curve visually represents the trade-off between true positive rates and false positive rates.

\subsubsection{Latency and Throughput}

These metrics measure the time taken to process transactions and the number of transactions processed per unit of time, respectively. They are essential for evaluating the efficiency and scalability of the proposed Blockchain and DLT solutions in handling large-scale data sharing scenarios.

\subsubsection{Security and Privacy Metrics}

Metrics such as data leakage rate, cryptographic strength, and resilience to attacks can be used to evaluate the security and privacy aspects of Blockchain and DLT solutions. Researchers should focus on minimizing data leakage rates while maintaining strong cryptographic protection and robustness against various attacks.

The choice of the most appropriate metric for evaluating the accuracy of Blockchain and DLT solutions for intelligent sharing depends on the specific requirements of the application domain. For instance, in a financial transaction system, minimizing false positives and negatives might be crucial, making F1 and F2 scores more relevant. In contrast, for a healthcare data sharing application, maintaining high data security and privacy might be a higher priority, necessitating the use of security and privacy metrics.

Implementing and evaluating Blockchain and DLT solutions for intelligent sharing requires a comprehensive experimental setup that addresses network architecture, security and privacy, and scalability and performance. Researchers can utilize various datasets, such as financial, healthcare, or IoT data, depending on the target application domain. To measure the accuracy of the proposed solutions, several metrics can be employed, including classification rates, F1 and F2 scores, AUC and ROC, and security and privacy metrics. The selection of the most appropriate metric should be guided by the specific requirements of the application domain.

\subsubsection{Future Directions and Challenges}
As the adoption of Blockchain and DLT for intelligent sharing continues to grow, several challenges and future directions emerge that warrant further research and development.

\textbf{Interoperability}

With the increasing number of Blockchain and DLT platforms and applications, interoperability between different systems becomes crucial for seamless and efficient data sharing. Future research should focus on developing standardized protocols and interfaces to enable communication between different Blockchain and DLT networks.

\textbf{Energy Efficiency}

Current consensus mechanisms, such as Proof of Work, are known for their high energy consumption, which has raised environmental concerns. Researchers should explore more energy-efficient consensus algorithms, such as Proof of Stake or novel alternatives, to address this challenge.

\textbf{Data Privacy Regulations}

Compliance with evolving data privacy regulations, such as the General Data Protection Regulation (GDPR), is essential for the widespread adoption of Blockchain and DLT solutions for intelligent sharing. Future research should investigate methods for ensuring compliance with such regulations while maintaining the benefits of decentralization and security.

\textbf{Adoption in Emerging Technologies}

Blockchain and DLT can play a significant role in securing emerging technologies such as the Internet of Things (IoT), 5G/6G networks, and edge computing. Future research should explore the integration of Blockchain and DLT solutions in these domains to enable secure and efficient data sharing.

\section{Conclusion}
In conclusion, this comprehensive review has addressed the key questions and topics related to intelligence sharing and blockchain-based intelligence sharing. We have examined the definition, objectives, benefits, challenges, and potential solutions associated with intelligence sharing, as well as the fundamentals of blockchain and distributed ledger technology.

Our analysis has shown that intelligence sharing is essential for enhancing security and mitigating risks associated with cyber attacks and other security threats. We have also identified the potential benefits of using blockchain and DLT for security and intelligence sharing, as well as the challenges and risks associated with their implementation. Furthermore, we have discussed the importance of a National Cybersecurity Strategy for addressing cybersecurity risks and examined the curricular ramifications of intelligence sharing.

Overall, our review suggests that blockchain and DLT offer promising solutions for enhancing security and intelligence sharing. However, further research is needed to evaluate the accuracy and performance of these solutions and to address the associated challenges and risks. Additionally, the integration of specific skills and knowledge related to intelligence sharing into existing curricula is crucial for preparing students and professionals to effectively address the challenges associated with intelligence sharing in the future.

% or


% \appendices

% % you can choose not to have a title for an appendix
% % if you want by leaving the argument blank
% \section{}

% \section*{Acknowledgment}

\ifCLASSOPTIONcaptionsoff
  \newpage
\fi

\end{document}